\documentclass[twocolumn,superscriptaddress,showpacs,preprintnumbers,amsmath,amssymb]{revtex4}
\usepackage[dvips]{color}
\usepackage{graphicx}
\usepackage{dcolumn}
\usepackage{bm}

\usepackage{amsmath,amssymb}

\usepackage{comment}

\begin{document}

\title{Fano resonance through Higgs bound states in tunneling of Nambu-Goldstone modes
}

\author{Takeru Nakayama}
\affiliation{Institute for Solid State Physics, University of Tokyo, 
5-1-5 Kashiwanoha, Kashiwa, Chiba 277-8581, Japan}

\author{Ippei Danshita}
\affiliation{Yukawa Institute for Theoretical Physics, Kyoto University, Kitashirakawa Oiwakecho, Sakyo-ku, Kyoto 606-8502, Japan}
\affiliation{Computational Condensed Matter Physics Laboratory, RIKEN, 2-1 Hirosawa, Wako, Saitama 351-0198, Japan}

\author{Tetsuro Nikuni}
\affiliation{Department of Physics, Faculty of Science, Tokyo University
of Science, 1-3 Kagurazaka, Shinjuku-ku, Tokyo 162-8601, Japan}

\author{Shunji Tsuchiya}
\affiliation{Center for General Education, Tohoku Institute of Technology, 35-1 YagiyamaKasumi-cho, Taihaku-ku, Sendai, Miyagi 982-8577, Japan}

\date{\today}

\begin{abstract}
We study collective modes of superfluid Bose gases in optical lattices combined with potential barriers. We assume that the system is in the vicinity of the quantum phase transition to a Mott insulator at a commensurate filling, where emergent particle-hole symmetry gives rise to two types of collective mode, namely a gapless Nambu-Goldstone (NG) phase mode and a gapful Higgs amplitude mode. We consider two kinds of potential barrier: One does not break the particle-hole symmetry while the other does. In the presence of the former barrier, we find Higgs bound states that have binding energies lower than the bulk Higgs gap and are localized around the barrier. 
We analyze tunneling properties of the NG mode incident to both barriers to show that the latter barrier couples the Higgs bound states with the NG mode, leading to Fano resonance mediated by the bound states.
Thanks to the universality of the underlying field theory, it is expected that Higgs bound states may be present also in other condensed matter systems with a particle-hole symmetry and spontaneous breaking of a continuous symmetry, such as quantum dimer antiferromagnets, superconductors, and charge-density-wave materials.
\end{abstract}

\maketitle
\section{Introduction}
\label{sec:Introduction}
The concept of elementary excitation is central to understanding various properties of quantum many-body systems, such as thermodynamics, transport, nonequilibrium dynamics, superfluidity, and phase transitions. It is of fundamental importance in modern condensed matter physics. In recent years, of particular interest are massive (or gapful) Higgs modes of systems with spontaneous breaking of a continuous symmetry that correspond to amplitude fluctuations of the order parameter, due to their ubiquity in many condensed-matter systems~\cite{volovik-14, pekker-14}. Examples include superconductors ${\rm NbSe}_2$~\cite{sooryakumar-80, littlewood-81, measson-14} and ${\rm Nb}_{1-x}{\rm Ti}_{x}{\rm N}$~\cite{matsunaga-13, matsunaga-14,sherman-15}, charge-density-wave (CDW) materials ${\rm K}_{0.3}{\rm MoO}_3$~\cite{demsar-99, schaefer-14} and ${\rm TbTe}_3$~\cite{yusupov-10, mertelj-13}, quantum dimer antiferromagnets ${\rm TlCuCl}_3$~\cite{ruegg-08} and ${\rm KCuCl}_3$~\cite{kuroe-12}, superfluid (SF) $^{3}{\rm He}$-B~\cite{avenel-80, collett-13}, and SF Bose gases in optical lattices~\cite{bissbort-11, endres-12}. Moreover, Higgs modes have attracted interest because of their close analogy with the Higgs boson in elementary particle physics.

All the Higgs modes that have been studied thus far are delocalized states in the entire system. In this paper, we study collective modes of SF Bose gases in optical lattices in the presence of potential barriers to predict bound states of Higgs mode that are localized around the barriers. Assuming the vicinity of the quantum phase transition to a Mott insulator (MI) at a commensurate filling, in which the system is nearly particle-hole symmetric~\cite{fisher-89, sachdev-11}, we analyze effects of potential barriers on the Higgs modes within the fourth order Ginzburg-Landau (GL) theory. We first consider a repulsive potential barrier that is created by locally reduced hopping amplitude and does not break the particle-hole symmetry. Near the barrier,
the static value of the superfluid order parameter locally diminishes. We show that the diminishing order parameter combined with the repulsive barrier constitutes a double well potential for the Higgs modes, thus leading to the formation of Higgs bound states.
Their binding energies are found to be lower than the energy gap of the Higgs mode in bulk which we call the ``bulk Higgs gap.''
We analytically obtain the energy and wave function of the Higgs bound states. 

Elementary excitations localized around edges or defects often play a crucial role in determining physical properties of the systems, especially transport properties, as is the case in the Andreev bound states in generic superconductors~\cite{andreev-64, furusaki-91}, the Dirac fermions in three-dimensional (3D) topological insulators~\cite{hasan-10}, and the Majorana fermions in topological superconductors~\cite{kitaev-01, fu-08}. For instance, a Josephson supercurrent flows through the Andreev bound states and the Dirac fermions carry edge currents in 3D topological insulators. We show that the presence of the Higgs bound states significantly affects the transport of gapless Nambu-Goldstone (NG) modes that correspond to phase fluctuations of the order parameter when there exists a potential barrier that is created by inhomogeneous chemical potential and breaks the particle-hole symmetry. More specifically, we consider a tunneling problem of the NG modes across the potential barriers and find Fano resonance~\cite{fano-61} of the NG modes mediated by the Higgs bound states. Existence of the Higgs bound states may be demonstrated through measurement of an asymmetric peak in the transmission probability characteristic to the Fano resonance.

The remainder of this paper is organized as follows. In Sec.~\ref{sec:Model}, we introduce the Bose-Hubbard (BH) model that describes Bose gases in optical lattices, and present a brief review of important properties of the model with an emphasis on the Higgs and NG modes of the SF phase. 
In Sec.~\ref{sec:TDGL}, we review the description of the Higgs and NG modes in a homogeneous system based on the GL theory.
In Sec.~\ref{sec:potential}, we explain a way to create potential barriers in the chemical potential and the hopping amplitude by controlling external fields and develop the GL theory to include the effects of the barriers. In Sec.~\ref{sec:HiggsBS}, we analyze the collective modes in the presence of the hopping barrier and reveal the emergence of Higgs bound states localized around the barrier. In Sec.~\ref{sec:Tunneling}, we solve a tunneling problem of the NG mode scattered by the two types of potential barrier and show that the Higgs bound states induce Fano resonance of the NG mode.
In Sec.~\ref{sec:QFT}, we construct a quantum field theoretical formulation of the collective modes in the presence of the potential that breaks the particle-hole symmetry.
The results are summarized in Sec.~\ref{sec:Conclusion}.

\begin{figure}
   \includegraphics[width=\linewidth]{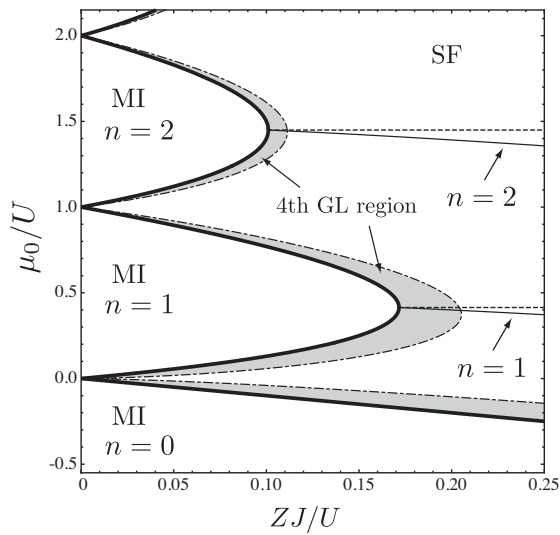}
   \caption{Ground-state phase diagram of the Bose-Hubbard model in a homogeneous system. 
The phase boundaries (thick solid lines) and the line of integer filling factors (thin solid lines) are computed by means of the Gutzwiller mean-field approximation. The dashed lines represent the particle-hole symmetric lines ($K_0=0$), which are obtained from the analytical expression for $K_0$ given in Appendix~\ref{sec:dev_tdgl}. The dashed-dotted line represents the contour of $\left|\psi\right|^2a^d=0.25$ and the gray shaded area roughly marks the region where the fourth-order GL theory is validated for describing the SF state.  }
 \label{fig:souzu}
\end{figure}

\section{Bose-Hubbard Model}
\label{sec:Model}
We consider cold bosonic atoms in a hypercubic optical lattice. We assume a sufficiently deep lattice so the system is well described by the tight-binding BH model
\begin{eqnarray}
\mathcal H=
-\sum_{{\bm i},{\bm j}}J_{{\bm i},{\bm j}}b_{\bm i}^\dagger b_{\bm j}
-\sum_{\bm i} \mu_{\bm i} b_{\bm i}^\dagger b_{\bm i}
+\frac{U}{2}\sum_{\bm i} b_{\bm i}^\dagger b_{\bm i}^\dagger b_{\bm i} b_{\bm i}.
\label{eq:BH}
\end{eqnarray}
The vector ${\bm i}\equiv \sum_{\alpha=1}^{d}i_{\alpha}{\bm e}_{\alpha}$ denotes the site index, where $i_{\alpha}$ is an integer, $d$ the spatial dimension, and ${\bm e}_{\alpha}$ a unit vector in direction $\alpha$. For instance, the directions $\alpha=1,2,$ and $3$ denote the $x,y,$ and $z$ directions, respectively. $b_{\bm i}^\dagger$ ($b_{\bm i}$) is a creation (annihilation) operator of bosons at site ${\bm i}$, 
and $U>0$ the on-site repulsive interaction. The local chemical potential,
\begin{eqnarray}
\mu_{\bm i}\equiv \mu_0 - V_{\bm i},
\label{eq:mui}
\end{eqnarray}
consists of the homogeneous contribution $\mu_0$ and the site-dependent external potential $V_{\bm i}$.  The hopping matrix element $J_{{\bm i},{\bm j}}$ is assumed to be finite only for nearest-neighboring sites, i.e., 
\begin{eqnarray}
J_{{\bm i},{\bm j}}
=\sum_{\alpha}
\left(J^{(\alpha)}_{\bm j} \delta_{{\bm i},{\bm j}+{\bm e}_{\alpha}}
+ J^{(\alpha)}_{{\bm j}-{\bm e}_{\alpha}} \delta_{{\bm i},{\bm j}-{\bm e}_{\alpha}}\right),
\end{eqnarray}
where $J^{(\alpha)}_{\bm j}$ means the hopping amplitude between sites ${\bm j}$ and ${\bm j}+{\bm e}_{\alpha}$.
We set $\hbar=1$ throughout the paper. 
Properties of ground states and low-lying elementary excitations of the
BH model in a homogeneous system ($J^{(\alpha)}_{\bm i}=J$, $V_{\bm
i}=0$) have been extensively studied and well
understood~\cite{krutitsky-15}. While we aim to reveal novel effects
caused by local potential barriers in an inhomogeneous system, in this and next sections we briefly review the properties of the homogeneous BH model in order to clarify the problem addressed in this paper.
\par
In Fig.~\ref{fig:souzu}, we show the ground-state phase diagram in the ($ZJ/U$,$\mu_0/U$)-plane obtained by mean-field theories~\cite{sheshadri-93, oosten-01}, where $Z$ is the coordination number. 
It consists of two distinct phases: the MI phase and SF phase~\cite{fisher-89}. For large interaction energy ($ZJ\ll U$) at a commensurate filling, the system is in the MI phase where integer number of bosons localize in each lattice site to avoid the large energy cost of repulsive interaction. 
For large kinetic energy ($ZJ \gg U$), the system is in the SF phase, where bosons can move around and condense in the lowest energy state. The global U(1) symmetry is broken in the SF phase, while there is no broken symmetry in the MI phase. The quantum phase transition that involves spontaneous breaking of U(1) symmetry takes place at a certain value of $ZJ/U$. This ratio between kinetic and interaction energy can be arbitrarily controlled by tuning the laser intensity of the lattice potential in a single system. A signature of the quantum phase transition was observed in the drastic change of the interference pattern of an atomic cloud released from a trapping potential~\cite{greiner-02}.
\par
Elementary excitations in the MI phase correspond to excess particles or holes~\cite{oosten-01, elstner-99, konabe-06}. The excitation spectrum has energy gap due to finite energy cost for adding or subtracting one particle. 
The SF phase in the close vicinity of the tips of the Mott lobes possesses two excitations, namely the gapless NG mode and the gapful Higgs mode~\cite{altman-02, huber-07}. These excitations arise from the broken U(1) symmetry and approximate particle-hole symmetry. The former corresponds to phase fluctuations of the order parameter and the latter corresponds to amplitude fluctuations. The U(1) gauge symmetry of our system is not local but global, because the superfluid is not coupled with a dynamical gauge field. As a consequence, there is no Higgs mechanism such that the NG mode remains gapless. As the system becomes far apart from the tips of the Mott lobes, the energy gap of the gapful mode rapidly increases and the mode turns into a single-particle excitation. 
In the deep SF regime ($ZJ\gg U$), there remains only the gapless NG mode as low-lying excitations~\cite{smerzi-02}, which is often referred to as the Bogoliubov mode~\cite{bogoliubov-47}.

\section{NG and Higgs modes in the TDGL equation}
\label{sec:TDGL}
Since our focus in this paper is on NG and Higgs modes in the vicinity of the tips of the Mott lobes, we continue further review on these modes in this section.
Near the SF-MI transition point, it is reasonable to expand the action in terms of the SF order parameter $\psi$. The definition of $\psi$ is given in Appendix~\ref{sec:dev_tdgl}. As a result, an effective action $S_{\rm eff}(\{\psi\})$ as well as the classical equation of motion for $\psi$ can be obtained. Taking the low-energy and continuum limit, SF dynamics in the vicinity of the quantum critical point is governed by the time-dependent Ginzburg-Landau (TDGL) equation~\cite{fisher-89, sachdev-11},
\begin{eqnarray}
i K_0\frac{\partial \psi}{\partial t}-W_0\frac{\partial^2\psi}{\partial t^2}=\left(-\frac{\nabla^2}{2m_{\ast}}+r_0+u_0|\psi|^2\right)\psi.
\label{eq:TDGL}
\end{eqnarray}
Here, $\psi({\bm x},t)$ denotes the SF order parameter at the position ${\bm x}\equiv a{\bm i}$ and the time $t$,  $m_{\ast}\equiv1/(2Ja^2)$ the effective mass, and $a$ the lattice constant.
Analytical expressions for the coefficients $K_0$, $W_0$, $r_0$, and $u_0$ at zero temperature as functions of the original BH parameters $(ZJ,\mu_0, U)$ are given in Appendix~\ref{sec:dev_tdgl}.

In Fig.~\ref{fig:souzu}, the gray shaded areas indicate the parameter regions where the TDGL equation is approximately valid.
\par
\begin{figure}
   \includegraphics[width=\linewidth]{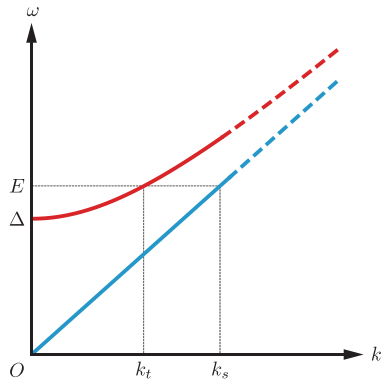}
   \caption{(Color online) Dispersion relations of the NG (lower blue) and Higgs (upper red) modes in
 Eqs.~(\ref{eq:phase}) and (\ref{eq:amplitude}). The NG mode
 has a gapless linear dispersion. The Higgs mode has energy gap $\Delta$.}  
 \label{fig:bunsan}
\end{figure}

When $K_0=0$, the TDGL equation is invariant with respect to the replacement $\psi \leftrightarrow \psi^{\ast}$, i.e., particle-hole symmetric. Moreover, this equation is mathematically a nonlinear Klein-Gordon equation that has the Lorentz invariance, and it is analogous to the relativistic field theory, where phase and amplitude modes are perfectly decoupled~\cite{higgs-64}. In Fig.~\ref{fig:souzu}, the lines of $K_0=0$ are plotted by the dashed lines that are parallel to the horizontal axis, and one sees that they are quite close to the lines of integer filling factors indicated by the thin solid lines.
\par
We specifically assume that the order parameter fluctuates from the equilibrium value $\psi_0=\sqrt{-r_0/u_0}$ as
\begin{eqnarray}
\psi=\psi_0+\delta\psi=\psi_0+\mathcal{U}(\bm{x})e^{-i\omega t}+\mathcal{V}^*(\bm{x})e^{i\omega^{\ast} t}.
\end{eqnarray}
Linearizing Eq.~(\ref{eq:TDGL}) with respect to fluctuations, we obtain
a set of equations,
\begin{eqnarray}
\left(-\frac{\nabla^2}{2m_*}+r_0+u_0\psi_0^2\right)S(\bm{x})&=&
\omega^2W_0S(\bm{x}), \label{eq:phase}\\
\left(-\frac{\nabla^2}{2m_*}+r_0+3u_0\psi_0^2\right)T(\bm{x})&=&
\omega^2W_0T(\bm{x}).\label{eq:amplitude}
\end{eqnarray}
Here, $S(\bm{x})\equiv\mathcal{U}(\bm{x})-\mathcal{V}(\bm{x})\propto \delta \theta(\bm{x})$ and $T(\bm{x})\equiv \mathcal{U}(\bm{x})+\mathcal{V}(\bm{x})\propto \delta n(\bm{x})$ correspond to phase and amplitude fluctuations of the order parameter, respectively, where $\psi=\sqrt{-r_0/u_0+\delta n(\bm x,t)}e^{i\delta\theta(\bm x,t)}$. Equations~(\ref{eq:phase}) and (\ref{eq:amplitude}) show that phase and
amplitude fluctuations are decoupled. Making Fourier transformation 
$(\mathcal{U}(\bm{x}),\mathcal{V}(\bm{x}))=(\mathcal{U}_{\bm{k}},\mathcal{V}_{\bm{k}})e^{i\bm{k}\cdot\bm{x}}$,
we obtain the dispersion relations for the NG and Higgs modes
\begin{eqnarray}
\begin{split}
\rm{NG} :&\quad \omega^2=c^2k^2, \label{NGdisp}\\
\rm{Higgs} :&\quad \omega^2=c^2k^2+\Delta^2. \label{Higgsdisp}
\end{split}
\end{eqnarray}
In Fig.~\ref{fig:bunsan}, we schematically show the dispersion relations.
The NG mode has a gapless linear dispersion, where $c=\sqrt{1/(2m_*W_0)}$ is the speed of sound. The Higgs mode 
has a finite gap $\Delta=\sqrt{-2r_0/W_0}$ at $k=0$. They are pure phase and amplitude modes for any $\bm k$. 
When $K_0\neq 0$, the particle-hole symmetry and the Lorentz invariance are broken so that the two modes are mixed. However, as long as $|K_0|$ is sufficiently small ($|K_0|\ll \sqrt{-W_0r_0}$), the basic property is robust, i.e., the phase and amplitude fluctuations dominate the gapless and gapful modes.
\par
Although Higgs modes apparently look long-lived within the linearized equations of motion (\ref{eq:phase}) and (\ref{eq:amplitude}), previous studies have dealt with higher order corrections with respect to the fluctuations and pointed out that the Higgs modes at $d<3$ are not necessarily well-defined because of strong quantum fluctuations allowing for decay of a Higgs mode into a pair of NG modes~\cite{altman-02, sachdev-99, podolsky-11, pollet-12, gazit-13, chen-13, rancon-14}. In the following, to avoid the subtlety at low dimensions, we focus on the case of $d=3$, where the use of TDGL equation is unambiguously justified (at least qualitatively) and Higgs modes are known to be long-lived.

\vspace{5mm}
\section{Effect of Potential Barriers in the TDGL equation}
\label{sec:potential}
%
\begin{figure}
   \includegraphics[width=\linewidth]{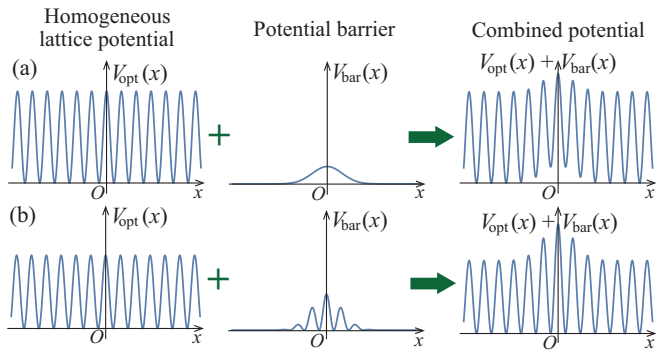}
   \caption{Schematic illustration of external potentials that create the inhomogeneous chemical potential $V_{i_1}$ (a) and hopping amplitude $J'_{i_1}$ (b). $V_{\rm opt}(x)$ and $V_{\rm bar}(x)$ mean the homogeneous optical-lattice potential and the potential barrier.}
\label{fig:potentials}
\end{figure}
In this section, we formulate effects of potential barriers in terms of the TDGL equation. We neglect a parabolic trapping potential for simplicity. Effects of a parabolic potential will be discussed at the end of Sec.~\ref{sec:HiggsBS}. We assume zero temperature in the following analyses, while our results regarding the Higgs bound states should be valid for realistic systems as long as the temperature is sufficiently low compared to their binding energy, which will be shown to be on the same order of magnitude as the bulk Higgs gap $\Delta$. Moreover, we assume that potential barriers are present only in the $x$ direction and that the system is homogeneous in the other directions except for the overall optical-lattice potential.
Let us show that external potentials can introduce the local modulation of the chemical potential $\mu_{\bm i}$ and the hopping amplitude $J_{\bm i}^{(\alpha)}$ in the BH model (\ref{eq:BH}). 
\par
Specifically, we propose imposing two different types of potential barrier in addition to the overall optical-lattice potential for controlling these parameters independently. 
First, the shift of the lattice potential with little change in the lattice
height can be realized by an optical dipole potential that leads to the
shift of the chemical potential $\mu_{\bm i}=\mu_0\to\mu_0-V_{i_1}$ in Eq.~(\ref{eq:BH}). This situation is schematically illustrated in Fig.~\ref{fig:potentials}(a).
Second, we consider an additional lattice potential in the Gaussian profile with the same
lattice spacing as that of the overall lattice potential as shown in Fig.~\ref{fig:potentials}(b).
The potential of this type can be created by focusing the optical-lattice laser into a narrow spatial region~\cite{miller-07} and spatially modulating the height of the lattice potential, leading to the inhomogeneous hopping amplitude,
\begin{eqnarray}
J_{\bm i}^{(\alpha)}= J+J'_{i_1}\delta_{\alpha,1}~.
\label{eq:Ji}
\end{eqnarray}
Since we regard $V_{i_1}$ and $J'_{i_1}$ as potential barriers, they are anticipated to vanish at $i_1 \to \pm \infty$.
Hence, $\mu_0$ and $J$ mean the equilibrium values far away from the potential barriers.
\par
The coefficients in the TDGL equation are modified by the potential barriers. 
We show approximate expressions of the coefficients in the lowest order of the perturbations $V_{i_1}$ and $J'_{i_1}$, taking the continuum limit $V_{i_1}\to V(x)$ and $J'_{i_1}\to J'(x)$.
See Appendix~\ref{sec:dev_tdgl} for a detailed derivation of the expressions.
We assume that $K_0=0$ such that there are independent NG and Higgs modes in the absence of the barriers. Here $K_0$, $W_0$, $r_0$, and $u_0$ denote the values of coefficients of the first-order time derivative term $K$, the second-order time derivative term $W$, the linear term $r$, and the cubic nonlinear term $u$ in the absence of the barriers. In the case that $K_0=0$, the shift of the chemical potential yields the leading contribution to $K$ as
\begin{eqnarray}
K \simeq -2W_0V(x)\equiv v_K(x).
\end{eqnarray}
This term breaks the particle-hole symmetry and locally couples phase and
amplitude fluctuations. In contrast, under the assumption that $V(x)\ll U$ one may ignore the contribution of $V(x)$ in $W$ and $u$ such that $W\simeq W_0$ and $u\simeq u_0$.
On the other hand, the local modulation of the hopping amplitude $J'(x)$ affects only $r$ as
\begin{eqnarray}
r \simeq r_0 - 2J'(x) \equiv r_0 + v_{r}(x).
\end{eqnarray}
$v_r(x)$ acts as a usual potential term that does not break particle-hole symmetry.
The resulting TDGL equation including the effects of the potential barriers is given by 
\begin{eqnarray}
iv_K\frac{\partial \psi}{\partial t} \!-\! W_0\frac{\partial^2 \psi}{\partial t^2}
\!=\!
\left(-\frac{\nabla^2}{2m_{\ast}}+r_0+v_r+u_0|\psi|^2\right)\psi.
\label{eq:tdgl_b}
\end{eqnarray}
In order to simplify the notation, we represent the variables in a dimensionless form,
\begin{eqnarray}
\begin{split}
\tilde{\psi}=\psi/(-r_0/u_0)^{1/2}, \,\, \tilde{t}=t(-r_0/W_0)^{1/2}, \,\, \tilde{x}=x/\xi, \\
\tilde{v}_r=-v_r/r_0, \,\,  \tilde{v}_K=v_K/(-r_0 W_0)^{1/2}.
\end{split}
\label{eq:Dless}
\end{eqnarray}
where $\xi \equiv (-m_{\ast}r_0)^{-1/2}$ is the healing length.
Hereafter, we omit the tilde and employ the
following TDGL equation in the dimensionless form,
\begin{equation}
iv_K\frac{\partial \psi}{\partial t}-\frac{\partial^2\psi}{\partial t^2}=\left(-\frac{\nabla^2}{2}-1+|\psi|^2+v_r\right)\psi.
\end{equation}
%
\begin{figure}
   \includegraphics[width=\linewidth]{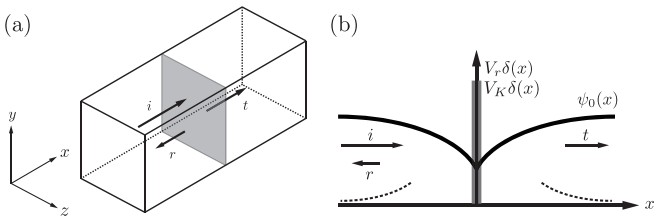}
   \caption{Schematic pictures of the geometry of the system (a) and tunneling NG mode in the $x$ direction through the $\delta$-function potential barriers $v_r(x) = V_r\delta(x)$ and $v_K(x) = V_K\delta(x)$ combined with the diminishing condensate $\psi_0(x)$ (b). The arrows mean plane waves of NG mode incident to the barrier from the left $(i)$, transmitting through the barrier $(t)$, and being reflected at the barrier $(r)$. The dashed line shows the profile $T(x)$ of the Higgs bound states localized around the barrier. }  
 \label{fig:schematic}
\end{figure}

We consider fluctuations of the order parameter $\psi({\bm x},t)$ around its static value $\psi_0({\bm x})$,
\begin{eqnarray}
\psi({\bm x},t)=\psi_0({\bm x})+\mathcal{U}({\bm x})e^{-i\omega t}+\mathcal{V}^*({\bm x})e^{i\omega^* t}.
\end{eqnarray}
The static order parameter $\psi_0({\bm x})$ satisfies the nonlinear equation that is identical to the
static Gross Pitaevskii (GP) equation~\cite{pitaevskii-61} as
\begin{eqnarray}
 \left(-\frac{\nabla^2}{2}-1+|\psi_0({\bm x})|^2+v_r(x)\right)\psi_0(\bm{x})=0~.
\label{eq:static_3d}
\end{eqnarray}
Phase and amplitude fluctuations $S(\bm{x})$and $T(\bm{x})$ obey the coupled equations
\begin{eqnarray}
\begin{split}
\left(-\frac{\nabla^2}{2}-1+|\psi_0({\bm x})|^2+v_r(x)\right)&S(\bm{x})\\
=\omega^2S(\bm{x})+&\omega v_K(x)T(\bm{x})~, \label{eq:S(x)_3d}
\end{split}
\\
\begin{split}
\left(-\frac{\nabla^2}{2}-1+3|\psi_0({\bm x})|^2+v_r(x)\right)&T(\bm{x})\\
=\omega^2T(\bm{x})+&\omega v_K(x)S(\bm{x})~. \label{eq:T(x)_3d}
\end{split}
\end{eqnarray}
The potential barrier $v_K(x)$ appears in the above equations in a peculiar manner: $v_K(x)$ is absent in
Eq.~(\ref{eq:static_3d}), so it does not affect $\psi_0$. Meanwhile, $v_K(x)$ in the coefficients of the 
frequency $\omega$ in Eqs.~(\ref{eq:S(x)}) and (\ref{eq:T(x)}) locally couples
$S(x)$ and $T(x)$ at the position of the potential barrier.
We will observe the crucial role played by this potential term in the
resonant tunneling of NG mode in Sec.~\ref{sec:Tunneling}.

\par
In the following analyses, we assume $\delta$-function potential barriers
and set $v_r(x)=V_r\delta(x)$ and $v_K(x)=V_K\delta(x)$ for simplicity.
This assumption is justified if the potential barrier spatially varies in
the order of lattice spacing that is much smaller than the healing
length $\xi$.
Since $\xi$ becomes very large in the quantum critical region near the
phase boundary with the MI phase, this assumption is reasonable when the
TDGL equation is valid. 
In the 3D geometry, the potential barriers take a sheetlike shape, as depicted in Fig.~\ref{fig:schematic}.

Since the system is assumed to be homogeneous in the $yz$ plane, the static order parameter in the ground state is independent of $y$ and $z$. Hence, Eq.~(\ref{eq:static_3d}) reduces to
\begin{eqnarray}
 \left(-\frac{1}{2}\frac{d^2}{dx^2}-1+|\psi_0(x)|^2+v_r(x)\right)\psi_0(x)=0~.
\label{eq:static}
\end{eqnarray}
Moreover, the fluctuations are simply described as plane waves in the $yz$ direction,
\begin{eqnarray}
S({\bm x})=S_{\rm 1D}(x)e^{i(k_{s,y} y + k_{s,z} z)},
\\
T({\bm x})=T_{\rm 1D}(x)e^{i(k_{t,y} y + k_{t,z} z)}.
\end{eqnarray}
In the following analyses, we assume that NG and Higgs modes propagate only in the $x$-direction, i.e., $k_{s,y}=k_{s,z}=k_{t,y}=k_{t,z}=0$. The Eqs.~(\ref{eq:S(x)_3d}) and (\ref{eq:T(x)_3d}) reduce to
\begin{eqnarray}
\begin{split}
\left(-\frac{1}{2}\frac{d^2}{dx^2}-1+|\psi_0(x)|^2+v_r(x)\right)&S_{\rm 1D}(x)\\
=\omega^2S_{\rm 1D}(x)+&\omega v_K(x)T_{\rm 1D}(x)~, \label{eq:S(x)}
\end{split}
\\
\begin{split}
\left(-\frac{1}{2}\frac{d^2}{dx^2}-1+3|\psi_0(x)|^2+v_r(x)\right)&T_{\rm 1D}(x)\\
=\omega^2T_{\rm 1D}(x)+&\omega v_K(x)S_{\rm 1D}(x)~. \label{eq:T(x)}
\end{split}
\end{eqnarray}
We henceforth rewrite $S_{\rm 1D}(x)$ and $T_{\rm 1D}(x)$ as $S(x)$ and $T(x)$ for brevity.

\section{Higgs bound states}
\label{sec:HiggsBS} 
To investigate localized bound states induced by $v_r(x)$, we assume $v_K(x)=0$ throughout this section.
The static solution under a $\delta$-function potential barrier $v_r(x)=V_r\delta(x)$ is given by \cite{kovrizhin-01}
\begin{eqnarray}
\psi_0(x)=\tanh\left(|x|+x_0\right)~,
\label{eq:staticpsi0}
\end{eqnarray}
where $x_0$ is determined by the boundary condition at $x=0$ 
\begin{eqnarray}
&\displaystyle \psi_0(-0)=\psi_0(+0), \label{bdcondp1}\\
&\displaystyle \left.\frac{d \psi_0}{d x}\right|_{+0}-\left.\frac{d \psi_0}{d
		x}\right|_{-0}=2V_r\psi_0(0),
\label{bdcondp2}
\end{eqnarray}
to give
\begin{equation}
\tanh(x_0)=-\frac{V_r}{2}+\sqrt{\frac{V_r^2}{4}+1}\equiv \eta.
\end{equation}
The amplitude of the static condensate at $x=0$, $\psi_0(0)=\eta$, monotonically decreases from $\eta(V_r=0)=1$ with
increasing $V_r$ and has the asymptotic form $\eta(V_r\to\infty)\sim 1/V_r$.
\par
As one can see from Eqs.~(\ref{eq:S(x)}) and (\ref{eq:T(x)}), the
diminishing $\psi_0(x)$ combined with the repulsive potential barrier
$v_r(x)$ constitutes a double-well potential for the collective modes. 
We demonstrate that Eq.~(\ref{eq:T(x)}) with the double-well potential allows bound-state solutions of
amplitude fluctuations below the bulk Higgs gap $\Delta$ that localize around the
potential well.
\par
Since $\psi_0(x)$ on the left (right) side of the barrier $v_r(x)$ is identical to the kink solution 
shifted by $x_0$ ($-x_0$), adopting the solutions of $T(x)$ on a static kink condensate in Appendix \ref{sec:kinksolution}, the solution of $T(x)$ for $\omega=\sqrt{2-\kappa_t^2/2}(<\Delta)$ reads
\begin{eqnarray}
&\displaystyle \scalebox{0.90}{$ T(x)=\left\{
\begin{array}{ll}
A\frac{3\psi_{0}^2 + 3\kappa_t \psi_{0} + \kappa_t^2 - 1}
{2+3\kappa_t + \kappa_t^2}e^{\kappa_tx}, 
& (x<0)
\\
\\
B\frac{3\psi_{0}^2 + 3\kappa_t \psi_{0} + \kappa_t^2 - 1}
{2+3\kappa_t + \kappa_t^2}e^{-\kappa_tx},
& (x>0)
\end{array} \right.$}. 
&\displaystyle 
\label{eq:HBS_sol}
\end{eqnarray}
The bound-state solutions of $T(x)$ satisfy the boundary condition at
$x=0$ as
\begin{eqnarray}
&\displaystyle T(-0)=T(+0), \label{bdcond1}\\
&\displaystyle \left.\frac{d T}{d x}\right|_{+0}-\left.\frac{d T}{d
		x}\right|_{-0}=2V_rT(0).
\label{bdcond2}
\end{eqnarray}
Remarkably, the above equations have two solutions: $A=B$ and $A=-B$. 
If $T(0)\neq 0$, Eq.~(\ref{bdcond1}) reduces to $A=B$, while if $T(0)=0$
Eq.~(\ref{bdcond2}) reduces to $A=-B$. The former corresponds to an even-parity solution and the latter an odd-parity one. 
We note that Eq.~(\ref{eq:S(x)}) has no unstable bound-state solutions with imaginary 
$\omega$.
\par
The difference between Eqs. (\ref{eq:S(x)}) and (\ref{eq:T(x)}) concerning to existence
of bound-state solutions indeed derives from the potential terms of
static condensate proportional to $|\psi_0|^2$. The deeper potential
well in Eq.~(\ref{eq:T(x)}) than that in Eq.~(\ref{eq:S(x)}) gives rise to the Higgs gap and
accommodates the localized bound states.
The emergence of the bound states of amplitude fluctuations in
the TDGL equation should be compared with the case of the GP equation
that has no bound states of amplitude as well as phase fluctuations.
\par
\begin{figure}
   \includegraphics[width=\linewidth]{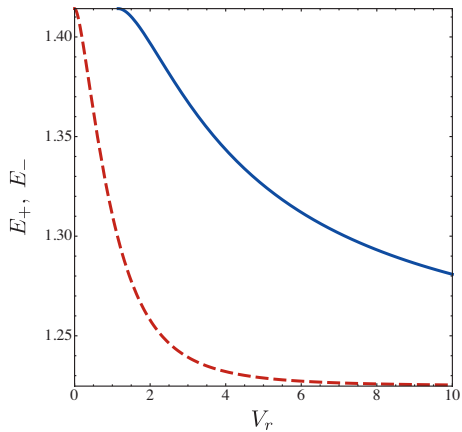}
   \caption{(Color online) Energy of the Higgs bound states with even parity ($E_+$) and odd parity ($E_-$) as functions of the potential strength $V_r$. The red dashed (blue solid) line shows $E_+$ ($E_-$). The vertical and horizontal axes are in units of $\sqrt{-r_0/W_0}$ and $-r_0\xi$, respectively.}
 \label{fig:E+E-}
\end{figure}
\begin{figure}[h]
   \includegraphics[width=\linewidth]{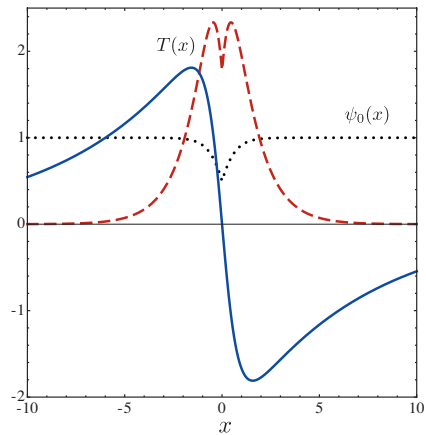}
   \caption{(Color online) Wave functions of the Higgs bound states $T(x)$ with parity even (red dashed line) and odd (blue solid line).  We set $V_r=1.5$
and $A=2+3\kappa_t +\kappa_t^2$. The black dotted line shows the static order parameter $\psi_0(x)$. The vertical and horizontal axes are in units of $\sqrt{-r_0/u_0}$ and $\xi$, respectively.}
 \label{fig:WF.eps}
\end{figure}
From Eqs.~(\ref{bdcond1}) and (\ref{bdcond2}), the even-parity bound state fulfills the condition
\begin{eqnarray}
c_1+V_rc_2=0~,
\label{eq:evenparity}
\end{eqnarray}
where 
\begin{eqnarray}
\begin{split}
c_1&=\kappa_t^3+3\eta\kappa_t^2+(6\eta^2-4)\kappa_t+6\eta(\eta^2-1), \\
c_2&=\kappa_t^2+3\eta\kappa_t+3\eta^2-1. 
\end{split}
\end{eqnarray}
Equation~(\ref{eq:evenparity}) has a single bound-state solution $\kappa_+$. 
Figure~\ref{fig:E+E-} shows the binding energy
$E_+=\sqrt{2-\kappa_+^2/2}$ as a function of $V_r$. $E_+(V_r)$ becomes the Higgs gap
$E_+\to\sqrt{2}$ when $V_r\to 0$. The bound state reduces to the odd-parity 
solution localized around a kink (see Appendix \ref{sec:kinksolution}): $E_+\to\sqrt{3/2}$ as
$V_r\to\infty$. In this limit, the bound state can be also considered as an edge
state that is localized at the boundary where condensate vanishes.
\par
The odd parity solution satisfies $c_2=0$. We thus obtain 
\begin{equation}
\kappa_t=\frac{1}{2}\left(-3\eta+\sqrt{4-3\eta^2}\right)\equiv\kappa_-~.
\end{equation}
The energy of the odd parity solution is given by $E_-=\sqrt{2-\kappa_-^2/2}$.
The odd parity bound state appears if the potential is large enough such that $V_r>2/\sqrt{3}$.
It also reduces to the odd-parity solution on a kink (see Appendix \ref{sec:kinksolution}): $E_-\to\sqrt{3/2}$ as $V_r\to\infty$. 
The odd-parity bound state has higher energy than the even parity one
($E_+<E_-$), as shown in Fig.~\ref{fig:E+E-}.
\par
Figure~\ref{fig:WF.eps} shows the even- and odd-parity bound states of $T(x)$. We propose the existence of such bound states of amplitude fluctuations below the Higgs gap and call them {\it Higgs bound states}. So far, the main focus of the study of localized excitations in condensed-matter systems have been on single-particle excitations, including Andreev bound states in superconductors~\cite{andreev-64} and edge states in quantum Hall systems~\cite{buttiker-88} and topological insulators~\cite{kane-05} or collective density modes such as ripplons~\cite{landaulifshitz-87} and Kelvin modes~\cite{thomson-1879} in SF systems. Hitherto, Higgs bound states as localized amplitude modes have never been found. Since the Higgs bound states are low-lying excitations, they may play a major role in various aspects of superfluid Bose gases in optical lattices at low temperatures. Moreover, given the fact that the presence of Higgs amplitude modes is a common feature among systems described effectively by a relativistic $O(N)$ field theory with $N\ge 2$~\cite{sachdev-11}, Higgs bound states are also expected to exist in other various systems involving approximate particle-hole symmetry and spontaneous breaking of a continuous symmetry, such as superconductors, CDW materials, and magnetic materials.

\begin{figure}[t]
   \includegraphics[scale=0.5]{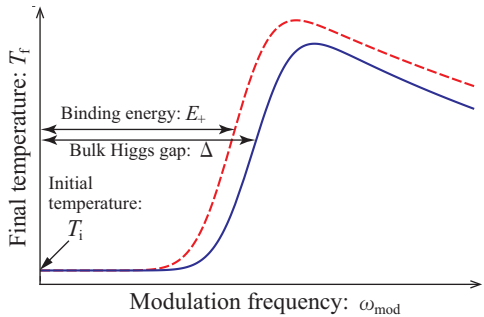}
   \caption{(Color online) Sketch of the temperature response to the temporal modulation of the lattice depth with frequency $\omega_{\rm mod}$. The blue solid and red dashed lines represent the final temperature $T_{\rm f}$ after the lattice modulation in the absence and presence of the hopping barrier potential, respectively.}
 \label{fig:trap}
\end{figure}

Let us discuss how to detect the Higgs bound states in cold-atom experiment. In the above calculations, we have ignored a global parabolic potential that confines atoms. Such a situation can be realized by combining an optical-lattice potential with a box-shaped confining potential, which has been achieved in recent experiment~\cite{gaunt-13}. To induce the Higgs bound states, one needs to add to the system the hopping barrier potential depicted in Fig.~\ref{fig:potentials}(a). Since the amplitude fluctuation directly responds to the temporal modulation of the lattice depth~\cite{huber-07}, it is expected that the binding energy of the even-parity Higgs bound state can be measured in the lattice-modulation spectroscopy as a sharp peak when the temperature is sufficiently low compared to the binding energy. Recall that the binding energy is well below the bulk Higgs gap, and therefore the Higgs bound state is the lowest-energy state that responds to the lattice modulation within the linear response regime.

Nevertheless, since a parabolic confining potential is used in most of the current experiments, it is important to mention its effects on detection of the Higgs bound states. In a superfluid gas confined in a parabolic potential, the local chemical potential spatially changes such that the particle-hole symmetry is present only locally in the close vicinity of the commensurate filling regions. This certainly makes the detection of Higgs modes in a parabolic trap more difficult than in a box-shaped potential. However, in recent experiment, Endres {\it et al.}~have successfully measured the bulk Higgs gap in the presence of a parabolic potential by using the lattice-modulation spectroscopy~\cite{endres-12}. Specifically, they measured the temperature rise after the lattice modulation of several cycles as a function of the modulation frequency. As sketched by the blue solid line in Fig.~\ref{fig:trap}, a sharp peak structure expected for a homogeneous system disappears in the measured temperature response due to the effect of the parabolic potential, and instead there is a response over a broad range of frequency. The onset of the spectral response was interpreted as the bulk Higgs gap. This interpretation is reasonable in the sense that there is no state responding to the lattice modulation below the bulk Higgs gap at the particle-hole symmetric point, and it is also supported by exact numerical analyses with use of quantum Monte Carlo simulations~\cite{pollet-12}.

When a hopping barrier potential is added to the system, the emergence of the even-parity Higgs bound state leads to the shift to the lower frequency side, as showned by the red dashed line in Fig.~\ref{fig:trap}. Notice that the binding energy monotonically increases when the system moves away from the particle-hole symmetric point (see Appendix~\ref{sec:spectral}), meaning that the onset frequency corresponds to the binding energy at the particle-hole symmetric point. We suggest that detecting the frequency shift will serve as an experimental signature of the Higgs bound state in the system confined in a parabolic potential.

%
%
\section{Fano resonance of Tunneling NG mode}
\label{sec:Tunneling}
In this section, we study scattering of NG mode in the presence of the
potential barriers $v_r(x)$ and $v_K(x)$.
We assume that NG mode with energy $E$ is injected from the left
$x\to-\infty$ as shown in Fig.~\ref{fig:schematic}(b). 
The solutions of Eqs.~(\ref{eq:S(x)}) and (\ref{eq:T(x)}) can be written
in a linear combination of the scattering states on a static kink
condensate in Eqs.~(\ref{eq:solutionS}) and (\ref{eq:solutionT}) as 
\begin{widetext}
\begin{eqnarray}
&&S(x)=\left\{
\begin{array}{ll}
\frac{\psi_0 + i k_s}{1+ik_s}e^{i k_s x}
+r_{\rm ng}\frac{\psi_0-ik_s}{1-i k_s} e^{-i k_s x}, &(x<0),
\\
\\
t_{\rm ng} \frac{\psi_0-ik_s}{1-i k_x}e^{ik_sx}, &(x>0),
\end{array} \right.\label{eq:tunnelS(x)}\
, \\
\nonumber \\
&&T(x)=\left\{
\begin{array}{ll}
r_{\rm h}\frac{3\psi_0^2 - 3 i k_t \psi_0 - k_t^2 - 1}{2-3ik_t-k_t^2}e^{-ik_tx},
 & (x<0),
\\
\\
t_{\rm h}\frac{3\psi_0^2 - 3 i k_t \psi_0 - k_t^2 - 1}{2-3ik_t-k_t^2}e^{ik_tx},
 & (x>0),
\end{array} \right.,
\label{eq:tunnelT(x)}
\end{eqnarray}
\end{widetext}
where $k_s=\sqrt{2}E$ and $k_t=\sqrt{2E^2-4}$ (see Fig.~\ref{fig:bunsan}).
In Eq.~(\ref{eq:tunnelS(x)}), $S(x<0)$ consists of injected and
reflected waves, while $S(x>0)$ is a transmitted wave. Since $S(x)$ and $T(x)$ are coupled by the potential
$v_K$, amplitude fluctuations may be induced and emitted from the potential
barrier. Equation~(\ref{eq:tunnelT(x)}) thus corresponds to plane waves
of Higgs mode propagating outward from the barrier for $E>\Delta$.
If injected NG mode has lower energy than the bulk Higgs gap ($E<\Delta$), then
$k_t$ should be substituted by $i\kappa_t=i\sqrt{4-2E^2}$ in
Eq.~(\ref{eq:tunnelT(x)}) so that $T(x)$ exponentially decays at
$|x|\to\infty$. In the following, we restrict ourselves within the
latter case of $E<\Delta$.
\par
The asymptotic forms of Eqs.~(\ref{eq:tunnelS(x)}) and
(\ref{eq:tunnelT(x)}) far away from the potential barriers are given by
\begin{eqnarray}
&&S(x)\rightarrow
\left\{
\begin{array}{l}
e^{ik_sx} + r_{\rm ng}e^{-ik_sx}, \quad(x\to-\infty) \\
\\
t_{\rm ng} e^{ik_sx}, \quad(x\to\infty)
\end{array}
\right.,
\label{eq:asymptoticS}\\
\nonumber \\
&&T(x)\rightarrow
\left\{
\begin{array}{l}
r_{\rm h}e^{\kappa_t x}, \quad(x\to-\infty)\\
\\
t_{\rm h}e^{- \kappa_t x}, \quad(x\to\infty)
\end{array}
\right.
.
\label{eq:asymptoticT}
\end{eqnarray}
From the ratio of the amplitudes of reflected and transmitted waves with respect to
that of the incident wave, the reflection and transmission probabilities
of NG mode are defined as $\mathcal{R}\equiv |r_{\rm ng}|^2$ and
$\mathcal{T}\equiv |t_{\rm ng}|^2$, respectively. They satisfy the
conservation law $\mathcal{R}+\mathcal{T}=1$. We derive
the conservation law for NG and Higgs modes in Appendix~\ref{sec:conservation}.
\par
The coefficients $r_{\rm ng}$, $t_{\rm ng}$, $r_{\rm h}$, and $t_{\rm
h}$ are determined so as to satisfy the boundary conditions: 
\begin{eqnarray}
&\displaystyle S(-0)=S(+0),\label{eq:S(0)}\\ &T(-0)=T(+0),\label{eq:T(0)}\\
&\displaystyle \left.\frac{d S}{d x}\right|_{+0}-\left.\frac{d S}{d x}\right|_{-0}=2V_rS(0)-2EV_KT(0), \label{eq:S'(0)}\\
&\displaystyle \left.\frac{d T}{d x}\right|_{+0}-\left.\frac{d T}{d x}\right|_{-0}=2V_rT(0)-2EV_KS(0). \label{eq:T'(0)}
\end{eqnarray}
\par
\begin{figure}
\centerline{\includegraphics[width=\linewidth]{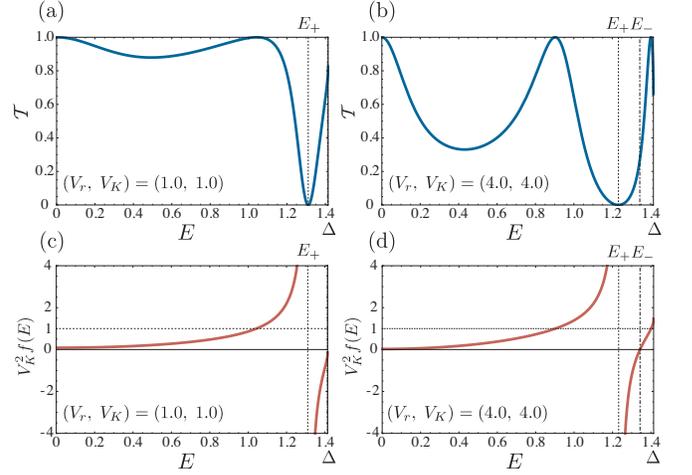}}
 \caption{(Color online) The upper panels represent the transmission coefficient $\mathcal{T}$ as a function of $E$ for $(V_r,V_K)=(1.0,1.0)$ (a) and $(4.0,4.0)$ (b).
The lower panels represent the scattering amplitude of NG mode $V_K^2f(E)$ as a function of $E$ for $(V_r,V_K)=(1.0,1.0)$ (c) and $(4.0, 4.0)$ (d). The dotted and dash-dotted lines represent the energy of the Higgs bound states with parity even $(E_+)$ and odd $(E_-)$, respectively. The horizontal axis is in the unit of $\sqrt{-r_0/W_0}$.}
 \label{fig:Tandf.eps}
\end{figure}

The transmission probability of NG mode $\mathcal{T}(E)=|t_{\rm ng}|^2$ can be cast in the form
\begin{eqnarray}
&&\mathcal{T}(E)^{-1}=1+\frac{2E^2}{(2E^2+1)^2}V_{\rm eff}(E)^2~,\label{eq:transmission1}\\
&&V_{\rm eff}(E)= \left(1-V_K^2f(E)\right)V_r,\label{eq:effectiveVr}\\
&&f(E)=\frac{c_2}{c_1+V_rc_2}\left(\frac{2E^2+\eta^2}{2V_r}\right).\label{eq:f(E)}
\end{eqnarray}
Since the effect of $V_K$ appears as its square in Eq.~(\ref{eq:transmission1}), the transmission probability is independent of the sign of $V_K$.
Figure~\ref{fig:Tandf.eps} shows
$\mathcal{T}(E)$ and $V_K^2f(E)$ as functions of $E$. In
Figs.~\ref{fig:Tandf.eps}(a) and 8(b), $\mathcal{T}(E)$   
increases as $E$ decreases at low energy ($E\lesssim 0.5$), and it approaches unity at $E\to 0$.
In fact, Eq.~(\ref{eq:transmission1}) clearly shows the perfect transmission
of NG mode occurring in the low-energy limit, i.e., $\mathcal{T}\to 1$ at $E\to
0$, irrespective of the strength of the potential barriers
$V_r$ and $V_K$. This is well known as the anomalous tunneling of
Bogoliubov mode \cite{
kovrizhin-01,kagan-03,danshita-06,danshita-07,kato-08,tsuchiya-09,ohashi-08,watabe-11}.
The anomalous tunneling has been mainly discussed in the context of
weakly interacting Bose gases based on the GP equation.
Our results show that the NG mode in a strongly interacting Bose system
also exhibits the anomalous tunneling property.
Recently, it has been proposed that the anomalous tunneling is a universal
behavior of the NG mode in systems with a broken continuous symmetry~\cite{kato-12}. 
\par
Figures~\ref{fig:Tandf.eps}(a) and 8(b) also show a peculiar asymmetric peak:
$\mathcal{T}(E)$ is sharply enhanced after dropping to zero in the
vicinity of $E_+$ as $E$ decreases below the gap. 
This asymmetric peak is the main focus of the present paper.
\par
Equation (\ref{eq:effectiveVr}) shows that the interference between scattered
waves of NG mode in two processes, one directly scattered by the bare $V_r$ and the other one by $V_K$
as well as by $V_r$, renormalizes $V_r$ giving the effective
potential $V_{\rm eff}(E)$. Moreover, Eq.~(\ref{eq:f(E)}) shows that the second 
process involves resonant excitation of the Higgs bound state through
the scattering amplitude $f(E)$:
Expansion of the denominator in Eq.~(\ref{eq:f(E)}) around $E_+$ gives
\begin{eqnarray}
&&c_1+V_rc_2\simeq \alpha(E_+-E),\label{eq:expansionE+}\\
&&\alpha=\frac{2E_+}{\kappa_+}\left[3\kappa_+^2+2\left(2\eta+\frac{1}{\eta}\right)\kappa_+
			       +3\eta^2-1\right],
\end{eqnarray}
where $\kappa_+=\sqrt{4-2E_+^2}$. Thus, $f(E)$ has a pole and diverges at $E_+$, as shown in Figs.~\ref{fig:Tandf.eps}(c) and 8(d). 
If the interference is destructive, then $V_{\rm eff}(E)$
vanishes and perfect transmission of incident wave occurs when
$V_K^2f(E)=1$. On the other hand, precisely at the energy 
of the bound state ($E=E_+$), $V_{\rm eff}$ diverges due to the
resonance with the Higgs bound state and therefore incident wave is perfectly reflected. 
Thus, such interference of scattered waves of NG mode produces the asymmetric
peak in Figs.~\ref{fig:Tandf.eps}(a) and (8b).
\par
This phenomenon is a typical example of {\it Fano resonance} \cite{fano-61},
in which interference between a directly scattered wave within continuum
and a resonantly scattered wave involving excitation of bound states produces
asymmetric peaks of scattering cross-section or transmission
probability. 
The Fano resonance of the NG mode in the present case exhibits interesting features.
One remarkable feature is that the Higgs bound state is
resonantly coupled with the NG mode by the potential barrier of the first-order time-derivative
term that arises due to the broken particle-hole symmetry. This differs considerably 
from usual single-particle scatterings described by the Schr\"odinger
equation where scattering states and bound states are
coupled by proximity of wave functions through a potential barrier.
\par
In Eq.~(\ref{eq:effectiveVr}), the effect of $V_K$ vanishes and the
effective potential $V_{\rm eff}$ reduces to the bare potential $V_r$ at $E=E_-$,
because of $c_2(E_-)=0$ and $f(E_-)=0$. Thus, in contrast with the
even-parity bound state at $E_+$ that causes the resonance [$(E_+)=\pm\infty$], the odd parity bound state $E_-$ cancels
the effect of the potential barrier $V_K$, because the wave function of
the odd parity bound state has a node at the position of the
potential barrier $x=0$.
\par
\begin{figure}
\centerline{\includegraphics[width=\linewidth]{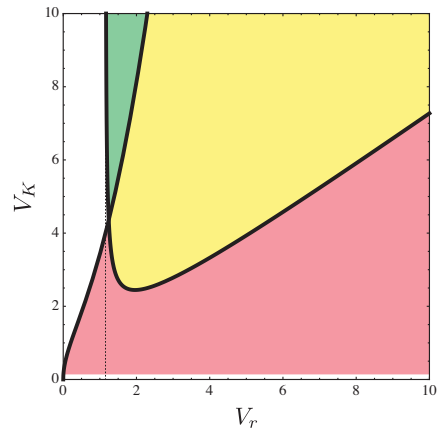}}
 \caption{(Color online) Phase diagram in terms of the perfect transmission of NG mode 
 in the $V_r-V_K$ plane. In the red (green) area, perfect transmission associated with the Higgs bound states occurs only once at energy of $0<E<E_+$ ($E_-<E<\Delta$). In the yellow area, Perfect transmission occurs twice with energies of $0<E<E_+$ and $E_-<E<\Delta$. The vertical and horizontal axes are in units of $\sqrt{-r_0 W_0}\xi$ and $-r_0\xi$.}
 \label{fig:souzu2.eps}
\end{figure}

If the odd-parity bound state exists ($V_r>2/\sqrt{3}$) and, furthermore, $V_K$ is
sufficiently large such that $V_K^2f(\Delta)>1$, then another perfect
transmission in the region $E_+<E<\Delta$ occurs when $V_K^2f(E)=1$ in addition to the one in $0<E<E_+$. 
Figure~\ref{fig:Tandf.eps}(b) shows the second perfect transmission in $E_+<E<\Delta$ for $(V_r,V_K)=(4.0,4.0)$.
The phase diagram in Fig.~\ref{fig:souzu2.eps} shows the parameter region where
perfect transmission occurs twice in the $V_r-V_K$ plane. 
\par
The observability of Higgs modes is a central issue in condensed-matter systems~\cite{pekker-14,varma-02}. 
Observation of Higgs modes as well as Higgs bound states is difficult with
standard techniques since they are not directly coupled with density or electromagnetic fields.
Few exceptions include observation in bosonic superfluids in optical lattices with temporal
modulation of the lattice potential~\cite{endres-12}, NbSe$_2$, which has
coexisting CDW and superconducting order, by Raman spectroscopy~\cite{sooryakumar-80}, and terahertz
transmission experiments in $s$-wave superconductors~\cite{matsunaga-13,matsunaga-14}.
Our results indicate that studying transport properties of NG mode could
be a possible platform for observation of Higgs bound states.
We propose detection of Higgs bound states in measuring the
transmission probability of the NG mode excited by Bragg
pulses~\cite{kozuma-99, stenger-99} through potential barriers~\cite{danshita-06}.
Since the asymmetric peak in the transmission probability of the NG mode is
characteristic to the Fano resonance coupled with the Higgs bound
states, detecting it provides with strong evidence for the existence of the Higgs bound states.

\section{Quantum field theory}
\label{sec:QFT}

In Sec.~\ref{sec:potential}, we have derived Eqs.~(\ref{eq:S(x)}) and (\ref{eq:T(x)}), which describe the collective modes of superfluid Bose gases in optical lattices, by linearizing the TDGL equation (\ref{eq:tdgl_b}) with respect to small fluctuations from a static value of the classical SF order-parameter field. Although such a derivation is simple and intuitive, the collective modes are treated as linear classical waves in its formalism such that one cannot compute higher-order corrections due to quantum fluctuations. In this section, we present a formulation of the collective modes based on quantum field theory, in which the collective modes are treated as quantized quasi-particles (elementary excitations). The quantum field theoretical formulation allows for inclusion of quantum corrections on the basis of the celebrated Green's function method. Notice that quantum field theory has been already established in the case that either the first- or second-order time-derivative term is present in the effective action of Eq.~(\ref{eq:Seff})~\cite{sachdev-11, popov-83}. Here we develop quantum field theory in the presence of both time-derivative terms in the action.

Let us start with the effective action of Eq.~(\ref{eq:Seff}) derived in Appendix A. While in the analyses of the previous sections we ignored the spatial dependence of the coefficients $W({\bm x})$ and $u({\bm x})$, which is only a small correction to the leading contribution, we here keep it as well as that of $K({\bm x})$ and $r({\bm x})$ in order for the formulation to be generic.
We split the field $\psi({\bm x},\tau)$ into its static value $\psi_0({\bm x})$ and fluctuations $\sigma({\bm x},\tau)$ and $\varphi({\bm x},\tau)$,
\begin{eqnarray}
\psi({\bm x},\tau) = \psi_0({\bm x}) + \sigma({\bm x},\tau) + i\varphi({\bm x},\tau),
\label{eq:sigmaphi}
\end{eqnarray}
where $\sigma$ and $\varphi$ are real variables.
Since the system has the global U(1) gauge invariance, we choose the gauge in which $\psi_0$ is real without loss of generality. In this gauge, $\sigma$ and $\varphi$ correspond to the amplitude and phase fluctuations as long as $\sigma, \varphi \ll \psi_0$. Substituting Eq.~(\ref{eq:sigmaphi}) into Eq.~(\ref{eq:Seff}), we expand the action in terms of the order of the fluctuations,
\begin{eqnarray}
S_{\rm eff} =  S_{0} +S_{1} + S_{2} + S_{3} + S_{4}.
\end{eqnarray}
The zeroth term $S_{0}$ is unimportant because it does not contain any dynamic degrees of freedom.
The linear term $S_{1}$ vanishes under the condition that $\psi_0$ satisfies Eq.~(\ref{eq:static_3d}).
The quadratic term is given by 
\begin{eqnarray}
S_{2} \!=\!\! \int \!\! d\tau d^d x \!\! \left[\sum_{\chi}\left(
\! W \! \left(\frac{\partial \chi}{\partial \tau}\right)^2
\!+ \! \chi \mathcal{H}_{\chi}\chi
\right)
\!+\! i2K\sigma\frac{\partial\varphi}{\partial\tau}
\right]\! ,
\label{eq:S2}
\end{eqnarray}
where $\chi \in \left\{\sigma, \varphi \right\}$ and
\begin{eqnarray}
\mathcal{H}_{\sigma} &=& -\frac{\nabla^2}{2m_{\ast}}+r+3u\psi_0^2, \\
\mathcal{H}_{\varphi} &=& -\frac{\nabla^2}{2m_{\ast}}+r+u\psi_0^2.
\end{eqnarray}
The higher-order terms are written as
\begin{eqnarray}
S_3 &=& \int d\tau d^d x \,
2 u \psi_0\left(
\sigma^3 + \sigma \varphi^2
 \right),
 \\
 S_4 &=& \int d\tau d^d x \, \frac{u}{2}(\sigma^2 + \varphi^2)^2.
\end{eqnarray}
In quantum field theory, it is common practice to diagonalize the quadratic term and treat the higher-order terms with a perturbation theory on the basis of the diagonalized quasi-particles. In the following, we will discuss how to diagonalize $S_2$ in the presence of both first- and second-order time-derivative terms.

In the absence of the second-order time-derivative term in the action ($W=0$), the amplitude and phase fluctuations are canonically conjugate variables. Since these conjugate variables are explicitly present in $S_2$ of Eq.~(\ref{eq:S2}), one can diagonalize $S_2$ by a linear transformation of $(\sigma, \varphi)$.
In contrast, when $W\neq 0$, the amplitude and phase fluctuations are no longer conjugate, and the conjugate variable to $\chi$, which corresponds to its ``momentum", does not explicitly exist in $S_2$ of Eq.~(\ref{eq:S2}). This means that $S_2$ of Eq.~(\ref{eq:S2}) cannot be diagonalized by a linear transformation of $(\sigma, \varphi)$ except for the special case of $K=0$. In order to introduce the conjugate variable as an independent variable to $\chi$ in the action, we perform a Stratonovich-Hubbard transformation, in which the conjugate variable $p_{\chi}$ is inserted into the action as an auxiliary field by using the Gaussian integral,
\begin{eqnarray}
\int \mathcal{D} p_{\chi} \exp \!\!\!\!\!&&\!\!\!\!\! \left( 
- \int d\tau d^d x
\frac{1}{4W}\left(p_{\chi} - i2W\frac{\partial \chi}{\partial \tau}\right)^2 
\right) 
\nonumber \\
&=& {\rm const.}
\label{eq:SH_trans}
\end{eqnarray}
Multiplying Eq.~(\ref{eq:SH_trans}) to the partition function, we rewrite the quadratic action as
\begin{eqnarray}
S_2 = \int d\tau d^d x&&\!\!\!\!\!\! \left[
\sum_{\chi}\left(-ip_{\chi}\frac{\partial \chi}{\partial \tau} + \frac{p_{\chi}^2}{4W}+\chi \mathcal{H}_{\chi}\chi \right)
\right.
\nonumber \\
&&\left.
+ i2K\sigma\frac{\partial\varphi}{\partial\tau}
\right].
\label{eq:S2_p}
\end{eqnarray}
As shown in the following, the quadratic action of the form of Eq.~(\ref{eq:S2_p}) can be diagonalized by a linear transformation of 
\begin{eqnarray}
{\boldsymbol \chi}({\bm x}, \tau) = 
\left[\sigma({\bm x}, \tau),
p_{\sigma}({\bm x}, \tau),
\varphi({\bm x}, \tau),
p_{\varphi}({\bm x}, \tau)\right]^{\bf t}.
\end{eqnarray}

We first perform the Fourier transform with respect to the imaginary time,
\begin{eqnarray}
{\boldsymbol \chi}({\bm x},\tau) = \frac{1}{\sqrt{\beta}}\sum_{\omega_n}e^{-i\omega_n\tau} 
\tilde{\boldsymbol \chi}({\bm x},\omega_n)
\label{eq:FT_chi}
\end{eqnarray}
where
\begin{eqnarray}
\!\!\!
\tilde{\boldsymbol \chi}({\bm x}, \omega_n) \!=\!
\left[\tilde{\sigma}({\bm x}, \omega_n),
\tilde{p}_{\sigma}({\bm x}, \omega_n),
\tilde{\varphi}({\bm x}, \omega_n),
\tilde{p}_{\varphi}({\bm x}, \omega_n)\right]^{\bf t} \!\! ,
\end{eqnarray}
and $\omega_n$ is the Matsubara frequency.
Substituting Eq.~(\ref{eq:FT_chi}) into Eq.~(\ref{eq:S2_p}), we obtain
\begin{eqnarray}
S_{2} = \sum_{\omega_n}\int d^d x\, 
\tilde{\boldsymbol \chi}^{\dagger}\hat{M}\tilde{\boldsymbol \chi},
\end{eqnarray}
where
\begin{eqnarray}
\hat{M}(\omega_n)=
\left[
\begin{array}{cccc}
H_{\sigma} & \frac{\omega_n}{2} & K\omega_n & 0 \\
-\frac{\omega_n}{2} & \frac{1}{4W} & 0 & 0 \\
-K\omega_n & 0 & H_{\varphi} & \frac{\omega_n}{2} \\
0 & 0 & -\frac{\omega_n}{2} & \frac{1}{4W} 
\end{array}
\right].
\label{eq:matM}
\end{eqnarray}
We next perform the linear transformation,
\begin{eqnarray}
\tilde{\boldsymbol \chi}({\bm x},\omega_n)
=\sum_{l}\hat{X}_{l}({\bm x}){\boldsymbol \alpha}_{l}(\omega_n),
\label{eq:cano}
\end{eqnarray}
where
\begin{eqnarray}
{\boldsymbol \alpha}_l(\omega_n)=\left[\alpha_l(\omega_n),
\alpha_l^{\ast}(-\omega_n)\right]^{\bf t}
\end{eqnarray}
and
\begin{eqnarray}
\hat{X}_l({\bm x})=
\left[
\begin{array}{cc}
\eta_{\sigma,l}({\bm x}) & \eta_{\sigma,l}^{\ast}({\bm x})
\\
\zeta_{\sigma,l}({\bm x}) & \zeta_{\sigma,l}^{\ast}({\bm x}) 
\\
\eta_{\varphi,l}({\bm x}) & \eta_{\varphi,l}^{\ast}({\bm x}) 
\\
\zeta_{\varphi,l}({\bm x}) & \zeta_{\varphi,l}^{\ast}({\bm x})
\end{array}
\right].
\label{eq:Xmat}
\end{eqnarray}
The coefficients for the transformation,
\begin{eqnarray}
{\bm y}_l({\bm x}) \equiv \left[\eta_{\sigma,l}({\bm x}),
\zeta_{\sigma,l}({\bm x}),
\eta_{\varphi,l}({\bm x}),
\zeta_{\varphi,l}({\bm x})
\right]^{\bf t},
\end{eqnarray}
satisfy the linear equation,
\begin{eqnarray}
\hat{M}(\omega_n \rightarrow -iE_l){\bm y}_l = 0,
\label{eq:bogolike}
\end{eqnarray}
where the index $l$ denotes the quantum number of the quasiparticles. 
While $E_l$ may be complex in general, we here assume that $E_l$ is real for all $l$. It is obvious that if the combination of ${\bm y}_l({\bm x})$ and $E_l$ is a solution of Eq.~(\ref{eq:bogolike}), that of ${\bm y}_l^{\ast}({\bm x})$ and $-E_l$ is also a solution. From Eq.~(\ref{eq:bogolike}), one can derive the orthogonality conditions,
\begin{eqnarray}
\int d^d x \left[
\sum_{\chi} \left(\eta_{\chi,l}^{\ast}\zeta_{\chi,l'}- \zeta_{\chi,l}^{\ast}\eta_{\chi,l'}\right)\right.
\nonumber \\
\left.
+ 2K\left( \eta_{\sigma,l}^{\ast}\eta_{\varphi,l'}-\eta_{\varphi,l}^{\ast}\eta_{\sigma,l'} \right)
\right]=C\delta_{l,l'},
\label{eq:ortho1}
\\
\int d^d x \left[
\sum_{\chi} \left(\eta_{\chi,l}\zeta_{\chi,l'}- \zeta_{\chi,l}\eta_{\chi,l'}\right)\right.
\nonumber \\
\left.
+ 2K\left( \eta_{\sigma,l}\eta_{\varphi,l'}-\eta_{\varphi,l}\eta_{\sigma,l'} \right)
\right]=0.
\label{eq:ortho2}
\end{eqnarray}
The requirement that the linear transformation of Eq.~(\ref{eq:cano}) has to be canonical determines the normalization constant in Eq.~(\ref{eq:ortho1}) as $C=-i$. Using Eqs.~(\ref{eq:ortho1}) and (\ref{eq:ortho2}), one can make the inverse transformation of Eq.~(\ref{eq:cano}),
\begin{eqnarray}
{\boldsymbol \alpha}_{l} = 
 \int d{\bf x} \, i\hat{\sigma}_z (\hat{X}_{l})^{\dagger}\hat{Q}\tilde{\bm \chi}
 \label{eq:inverse}
\end{eqnarray}
where
\begin{eqnarray}
\hat{\sigma}_{z}=
\left[
\begin{array}{cc}
1 & 0 \\
0 & -1
\end{array}
\right],
\end{eqnarray}
and
\begin{eqnarray}
\hat{Q} =
\left[
\begin{array}{cccc}
0 & 1 & 2K & 0 \\
-1 & 0 & 0 & 0 \\
-2K & 0 & 0 & 1 \\
0 & 0 & -1 & 0 
\end{array}
\right].
\label{eq:Qmat}
\end{eqnarray}

Equation (\ref{eq:bogolike}) indicates that once $\eta_{\chi,l}$ is obtained, $\zeta_{\chi,l}$ is determined trivially through the relation $\zeta_{\chi,l}=-2iWE_l\eta_{\chi,l}$.
Notice that with the relation $(\eta_{\sigma,l}({\bm x}),\eta_{\varphi,l}({\bm x}))\propto (T({\bm x}),-iS({\bm x}))$ one easily sees that Eq.~(\ref{eq:bogolike}) is equivalent to the linearized TDGL equations (\ref{eq:S(x)}) and (\ref{eq:T(x)}) while the normalization condition of Eq.~(\ref{eq:ortho1}) had not been imposed to the solutions of the latter equations, which are linear classical waves. This means that Eq.~(\ref{eq:ortho1}) may be interpreted as the quantization condition required for the collective modes to be regarded as quasiparticles.

Using Eqs.~(\ref{eq:bogolike}), (\ref{eq:ortho1}), and (\ref{eq:ortho2}), we diagonalize the quadratic action as
\begin{eqnarray}
S_{2} &=& \sum_{\omega_n,l,l'}\int d^d x\, 
{\boldsymbol \alpha}_{l'}^{\dagger}\hat{X}_{l'}^{\dagger}\hat{M}\hat{X}_{l}{\boldsymbol \alpha}_{l}
\nonumber \\
&=& \frac{1}{2}\sum_{\omega_n,l}\, 
{\boldsymbol \alpha}_{l}^{\dagger}
\left[
\begin{array}{cc}
-i\omega_n + E_l & 0 \\
0 & i\omega_n + E_l
\end{array}
\right]
{\boldsymbol \alpha}_{l}
\nonumber \\
&=& \sum_{\omega_n,l}
\alpha_{l}^{\ast}(\omega_n)\alpha_{l}(\omega_n)(-i\omega_n + E_l).
\end{eqnarray}
On the basis of the diagonalized quadratic action, the non-perturbative Green's function is given by
\begin{eqnarray}
G^{(0)}_l(i\omega_n) = -\langle \alpha_l(\omega_n)\alpha_l^{\ast}(\omega_n) \rangle_0
= \frac{1}{i\omega_n - E_l},
\end{eqnarray}
where the average $\langle \cdots \rangle_0$ is taken with the quadratic action $S_2$ as
\begin{eqnarray}
\langle O \rangle_0 = 
\frac{\int \mathcal{D}\alpha^{\ast} \mathcal{D}\alpha \,O\exp(-S_2)}
{\int \mathcal{D}\alpha^{\ast} \mathcal{D}\alpha \exp(-S_2)}.
\end{eqnarray}
While actual evaluations of quantum corrections to correlation functions are out of the scope of the present paper, standard quantum statistical mechanics tells that the diagonalized quadratic action serves as a foundation of the perturbative expansion of $S_3$ and $S_4$. More specifically, treating $S_3$ and $S_4$ as perturbation and using $G^{(0)}$ as the elementary piece of the perturbative expansion, one can systematically compute higher-order corrections to, for instance, the Green's function, 
\begin{eqnarray}
G_l(i\omega_n)=-\langle  \alpha_l(\omega_n)\alpha_l^{\ast}(\omega_n) \rangle 
\end{eqnarray}
through the Dyson's equation, 
\begin{eqnarray}
G_l(i\omega_n)
= \frac{1}{\left(G_l^{(0)}(i\omega_n)\right)^{-1} + \Sigma_l(i\omega_n)},
\end{eqnarray}
where $\Sigma_l(i\omega_n)$ is the self-energy and
\begin{eqnarray}
\langle O \rangle = 
\frac{\int \mathcal{D}\alpha^{\ast} \mathcal{D}\alpha \,O\exp(-S_{\rm eff})}
{\int \mathcal{D}\alpha^{\ast} \mathcal{D}\alpha \exp(-S_{\rm eff})}.
\end{eqnarray}
Moreover, as an application of the quantum field theoretical formulation, in Appendix~\ref{sec:spectral} we compute the spectral functions of the Higgs bound states, from which their decay rate can be evaluated, in the case where the first-order time-derivative term $K({\bm x})$ is finite.

\vspace{5mm}
\section{Conclusions}
\label{sec:Conclusion}
We have studied collective modes of SF Bose gases in optical
lattices in the presence of potential barriers. Assuming the system in
the vicinity of the quantum phase transition to the MI phase with
commensurate fillings, we derived the TDGL equation that includes the effect of external
potentials. 
We considered two types of potential barriers, one of which shifts the chemical potential and breaks the particle-hole symmetry,
while the other changes the hopping amplitude in the BH model, which does not break the particle-hole symmetry.
We found that introducing the former potential leads to the peculiar potential term coupled with the first-order
time-derivative of the SF order parameter in the TDGL equation.
In the presence of a potential barrier of the latter type, we have shown the
existence of Higgs bound states localized around the barrier below the
Higgs gap. We analyzed transport properties of the NG mode through the
potential barriers and found that the transmission probability of NG mode
exhibits a remarkable asymmetric peak that is characteristic to the Fano
resonance. We have shown that the Fano resonance of the NG mode involving
resonant excitation of Higgs bound states occurs due to the coupling of
phase and amplitude fluctuations induced by the potential barrier of the former type.
We proposed a possible way of detecting Higgs bound states in studying
transport properties of NG mode excited by Bragg pulses.

Moreover, we formulated quantum field theory for the collective modes of the system with both first- and second-order time derivative terms. This formulation will be crucial for future investigation of quantum corrections to the physics of the Higgs bound states analyzed within the quadratic approximation in this paper. It may be also interesting to apply the formulation to analyzing effects of breaking of the particle-hole symmetry on the decay rate of delocalized Higgs modes.

In this paper, we confined our discussions within the case in which the system has a single potential barrier of two different types. Given the fact that various systems, including disordered superconductors, Josephson junction arrays, and $^4$He absorbed in porous media, are effectively described by the BH model with random chemical potential and/or hopping amplitude \cite{fisher-89}, it may be desirable to extend our results to the case of random potential barriers. If potential barriers that change the local hopping amplitude distribute randomly or periodically over the system, then the Higgs bound states are expected to form energy bands below the Higgs gap. Such energy bands of Higgs bound states may be observable by measuring complex terahertz transmission \cite{sherman-15}.

\section*{Acknowledgments}
The authors thank S. Watabe for fruitful discussions. 
The authors also thank the Yukawa Institute for Theoretical Physics
(YITP) at Kyoto University, where this work was initiated during the
YITP workshop (YITP-W-14-02) on ``Higgs Modes in Condensed Matter and Quantum Gases.''. 
T. Nakayama thanks H. Tsunetsugu for useful comments.
T. Nakayama was supported by JSPS through Program for Leading Graduate Schools (MERIT).
The authors acknowledge Grants-in-Aid for Scientific Research from JSPS: Grants No.~25800228 (I. D), No.~25220711 (I. D), No.~25400419 (T. Nikuni), and No.~26800216 (S. T).
T.~Nakayama and I.~Danshita contributed equally to this work.

\vspace{5mm}
\appendix
\section{Derivation of time dependent Ginzburg-Landau equation}
\label{sec:dev_tdgl}
In this appendix, we present a detailed derivation of the TDGL equation (\ref{eq:tdgl_b}) that includes effects of inhomogeneous chemical potential $\mu_{\bm i}$ and hopping amplitude $J^{(\alpha)}_{\bm i}$ given by Eqs.~(\ref{eq:mui}) and (\ref{eq:Ji}). For this purpose, we describe the BH model of Eq.~(\ref{eq:BH}) in the imaginary-time path-integral representation as
\begin{eqnarray}
\Xi=\int \mathcal{D}b^*\mathcal{D}b\exp\left[-S_{\rm BH}(\{b_{\bm i}\})\right],
\label{eq:Xi1}
\end{eqnarray}
where $\Xi$ denotes the grand partition function and the Euclidian action is given by
\begin{eqnarray}
S_{\rm BH}(\{b_{\bm i}\}) = \int^{\frac{\beta}{2}}_{-\frac{\beta}{2}}  d\tau &&\!\!\!\!\!
\left[\sum_{\bm i} b_{\bm i}^*\left(\frac{\partial}{\partial \tau} - \mu_{\bm i} + \frac{U}{2}b_{\bm i}^{\ast}b_{\bm i}\right)b_{\bm i} \right. 
\nonumber \\ 
&&\Biggl. -\sum_{{\bm i},{\bm j}}J_{{\bm i},{\bm j}}b_{\bm i}^{\ast}b_{\bm j} \Biggr].
\end{eqnarray} 
We assume that $|V_{\bm i}|\ll U$ and $|J'_{i_1}|\ll J$. We follow the standard procedure used in previous studies~\cite{fisher-89, sachdev-11, yasukato-14} in most part of the derivation except for the treatment of the inhomogeneous hopping term.

We introduce the auxiliary field $\Psi_{\bm i}$ at site ${\bm i}$ that corresponds to the SF order parameter by Stratonovich-Hubbard transformation. This transformation makes use of the following Gaussian integral:
\begin{eqnarray}
\int \mathcal{D}\Psi^*\mathcal{D}\Psi &&\!\!\!\!\!\!\!\!\!\!   \exp\left(-\int d\tau \left(\vec{\Psi}^\dag-\vec{b}^\dag\hat{J}\right)\hat{J}^{-1}\left(\vec{\Psi}-\hat{J}\vec{b}\right)\right) \nonumber \\
&=& \rm{const.} 
\label{eq:SH}
\end{eqnarray}
Here, $\vec{b}\equiv(\left\{b_{\bm i}\right\})^T$ and $\vec{\Psi}\equiv(\left\{\Psi_{\bm i}\right\})^T$. $\hat{J}$ means the hopping matrix whose element is $J_{{\bm i},{\bm j}}$ and consists of the homogeneous part $\hat{J}_0$ and the inhomogeneous one $\hat{J}_{\rm bar}$ as
\begin{eqnarray}
\hat{J} = \hat{J}_{0} + \hat{J}_{\rm bar}.
\end{eqnarray}
where
\begin{eqnarray}
(\hat{J}_{0})_{{\bm i},{\bm j}} 
= \sum_{\alpha=1}^{d} J \left(\delta_{{\bm i},{\bm j}+{\bm e}_{\alpha}}
+\delta_{{\bm i},{\bm j}-{\bm e}_{\alpha}}\right)
\end{eqnarray}
and
\begin{eqnarray}
(\hat{J}_{\rm bar})_{{\bm i},{\bm j}} =  J'_{j_1} \delta_{{\bm i},{\bm j}+{\bm e}_{1}}
+J'_{j_1-1} \delta_{{\bm i},{\bm j}-{\bm e}_{1}}.
\end{eqnarray}
Multiplying Eq.~(\ref{eq:SH}) to Eq.~(\ref{eq:Xi1}), the grand partition function is rewritten as
\begin{eqnarray}
\Xi=\int \mathcal{D}b^*\mathcal{D}b\mathcal{D}\Psi^*\mathcal{D}\Psi\exp\left[-S(\{b_{\bm i}\},\{\Psi_{\bm i}\})\right],
\label{eq:Xi}
\end{eqnarray}
where
\begin{eqnarray}
&&\!\!\!\!\!\!\!\!\!\!\!\!
S(\{b_{\bm i}\},\{\Psi_{\bm i}\})= \int d\tau \vec{\Psi}^\dag \hat{J}^{-1} \vec{\Psi} +S_{\rm non}+S_{\rm{pert}}, \\
&&\!\!\!\!\!\!\!\!\!\!\!\!
S_{\rm non}=\int d\tau \sum_{\bm i} b_{\bm i}^* \left( \frac{\partial}{\partial \tau}-\mu_{\bm i}+\frac{U}{2}b_{\bm i}^*b_{\bm i}\right)b_{\bm i}, \\
&&\!\!\!\!\!\!\!\!\!\!\!\!
S_{\rm{pert}}=-\int d\tau (\vec{b}^\dag \vec{\Psi}+\vec{\Psi}^\dag \vec{b}).
\end{eqnarray}
Integrating out the bosonic fields $b_{\bm i}, {b}_{\bm i}^*$, the action is formally expressed as
\begin{eqnarray}
S=\beta F_{\rm non}+\int d\tau \vec{\Psi}^\dag \hat{J}^{-1} \vec{\Psi}-\ln\left<-S_{\rm pert}\right>_{\rm non},~
\label{eq:Sform}
\end{eqnarray}
where the average $\langle \cdots \rangle_{\rm non}$ is taken with the nonperturbative action $S_{\rm non}$ as
\begin{eqnarray}
\left<O\right>_{\rm non}=\frac{\int \mathcal{D}b^*\mathcal{D}b \exp\left(-S_{\rm non}\right)O}
{\int \mathcal{D}b^*\mathcal{D}b \exp\left(-S_{\rm non}\right)}~,
\end{eqnarray}
and $F_{\rm non}$ denotes the free energy of the MI state.
$S_{\rm non}$ contains only local terms and is already diagonalized with the filling factor $g$ as the good quantum number. Hence, the eigenstate of the system described by $S_{\rm non}$ is simply a Fock state $|g\rangle_{\bm i}$ and the eigenenergy is given by
\begin{eqnarray}
E_{g,{\bm i}}&=&-\mu_{\bm i} g+\frac{U}{2}g(g-1)~.
\end{eqnarray}
With these nonperturbative states and energies, it is straightforward to compute the average $\langle O \rangle_{\rm non}$, where the operator $O$ is supposed to consist of a product of $b_{\bm i}$ and $b_{\bm i}^{\ast}$.

Performing a cumulant expansion of the last term of Eq.~(\ref{eq:Sform}) at zero temperature up to the fourth order with respect to the fields $\Psi_{\bm i}$ and $\Psi_{\bm i}^{\ast}$, one obtains
\begin{eqnarray}
S(\{\Psi_{\bm i}\}) \!\!\!&=&\!\!\! \int d\tau \Biggl[ 
\sum_{{\bm i},{\bm j}} (\hat{J}^{-1})_{{\bm i},{\bm j}} \Psi_{\bm i}^{\ast}\Psi_{\bm j}
+ \sum_{\bm i}\biggl( \alpha^{(2)}_{\bm i}|\Psi_{\bm i}|^2\biggr.\Biggr. \nonumber \\
&&\!\!\!\!\!\!\!\!\!\!\!\!\!\!\!\!\!\!\!
\Biggl.\biggl. +\beta^{(2)}_{\bm i} \Psi_{\bm i}^{\ast} \frac{\partial\Psi_{\bm i}}{\partial \tau} 
+ \gamma^{(2)}_{\bm i}\left| \frac{\partial \Psi_{\bm i}}{\partial \tau} \right|^2
+ \alpha^{(4)}_{\bm i} |\Psi_{\bm i}|^4
\biggr)
\Biggr],
\label{eq:Smid}
\end{eqnarray}
where
\begin{eqnarray}
\alpha^{(2)}_{\bm i}&=&-\frac{g+1}{(E_{g+1,{\bm i}}-E_{g,{\bm i}})}
-\frac{g}{(E_{g-1,,{\bm i}}-E_{g,{\bm i}})}, \\
\alpha^{(4)}_{\bm i}&=&\left(\frac{g+1}{(E_{g+1,{\bm i}}-E_{g,{\bm i}})}
+\frac{g}{(E_{g-1,{\bm i}}-E_{g,{\bm i}})}\right)
\nonumber\\
&&\times\left(\frac{g+1}{(E_{g+1,{\bm i}}-E_{g,{\bm i}})^2}
+\frac{g}{(E_{g-1,{\bm i}}-E_{g,{\bm i}})^2}\right) 
\nonumber \\
&&-\frac{(g+1)(g+2)}{(E_{g+1,{\bm i}}-E_{g,{\bm i}})^2(E_{g+2,{\bm i}}-E_{g,{\bm i}})}
\nonumber \\
&&-\frac{g(g-1)}{(E_{g-1,{\bm i}}-E_{g,{\bm i}})^2(E_{g-2,{\bm i}}-E_{g,{\bm i}})}, \\
\beta^{(2)}_{\bm i}&=&\frac{g+1}{(E_{g+1,{\bm i}}-E_{g,{\bm i}})^2}
-\frac{g}{(E_{g-1,{\bm i}}-E_{g,{\bm i}})^2}, \\
\gamma^{(2)}_{\bm i}&=&\frac{g+1}{(E_{g+1,{\bm i}}-E_{g,{\bm i}})^3}+\frac{g}{(E_{g-1,{\bm i}}-E_{g,{\bm i}})^3}.
\end{eqnarray}
It is obvious that the coefficients $\alpha^{(2)}_{\bm i}$, $\beta^{(2)}_{\bm i}$, $\gamma^{(2)}_{\bm i}$, and $\alpha^{(4)}_{\bm i}$ reflect the inhomogeneity of the chemical potential while the first term in Eq.~(\ref{eq:Smid}) does that of the hopping. 
To clarify the latter effect, we transform the first term in Eq.~(\ref{eq:Smid}) under the assumption that $|J'_{i_1}|\ll J$,
\begin{eqnarray}
\int d\tau \sum_{{\bm i},{\bm j}} (\hat{J}^{-1})_{{\bm i},{\bm j}} \Psi_{\bm i}^{\ast}\Psi_{\bm j}
&=& \int d\tau\vec{\Psi}^{\dagger} \hat{J}^{-1}\vec{\Psi} \nonumber \\
&&\!\!\!\!\!\!\!\!\!\!\!\!\!\!\!\!\!\!\!\!\!\!\!\!\!\!\!\!\!\!\!\!\!\!\!\!\!\!\!\!\!\!\!\!\!\!\!\!\!\!
= \int d\tau\vec{\Psi}^{\dagger} \hat{J}_0^{-1}\left(1 + \hat{J}_0^{-1}\hat{J}_{\rm bar} \right)^{-1}\vec{\Psi} \nonumber \\
&&\!\!\!\!\!\!\!\!\!\!\!\!\!\!\!\!\!\!\!\!\!\!\!\!\!\!\!\!\!\!\!\!\!\!\!\!\!\!\!\!\!\!\!\!\!\!\!\!\!\!
\simeq\int d\tau \left(\vec{\Psi}^{\dagger} \hat{J}_0^{-1}\vec{\Psi} 
- \vec{\Psi}^{\dagger}\hat{J}_0^{-2}\hat{J}_{\rm bar} \vec{\Psi}\right).
\label{eq:Psihop}
\end{eqnarray}
Performing the Fourier transformation, the first and second terms in Eq.~(\ref{eq:Psihop}) are expressed as
\begin{eqnarray}
\int d\tau \!\!\!\!&&\!\!\!\!\!\!
\sum_{{\bm i},{\bm j}} \Psi_{\bm i}^{\ast} 
(\hat{J}_{0}^{-1})_{{\bm i},{\bm j}}\Psi_{\bm j}
= 
\sum_{{\bm k},\omega} |\tilde{\Psi}({\bm k},\omega)|^2\frac{1}{\varepsilon_{\bm k}}
\nonumber \\
&\simeq& \sum_{{\bm k},\omega} |\tilde{\Psi}({\bm k},\omega)|^2\frac{1}{ZJ}
\left(1-\frac{(ka)^2}{Z}\right),
\label{eq:Psi-1}
\end{eqnarray}
and
\begin{eqnarray}
 \int &&\!\!\!\!\!\!\!\!\!\!\! d\tau  \vec{\Psi}^{\dagger}\hat{J}_{0}^{-2}\hat{J}_{\rm bar}\vec{\Psi}
\nonumber \\
 &=& \sum_{\omega,{\bm k},{\bm k}'} 
 \tilde{\Psi}^{\ast}({\bm k},\omega)\tilde{\Psi}({\bm k}',\omega)
\frac{\tilde{J}'_{{\bm k}-{\bm k}'}}{\varepsilon_{\bm k}^2}(e^{ik_1a} + e^{-ik_1' a})
\nonumber \\
&\simeq& \sum_{\omega,{\bm k},{\bm k}'} 
\tilde{\Psi}^{\ast}({\bm k},\omega)\tilde{\Psi}({\bm k}',\omega)
\frac{2\tilde{J}'_{{\bm k}-{\bm k}'}}{(ZJ)^2},
\label{eq:Psi-2}
\end{eqnarray}
where
\begin{eqnarray}
\Psi_{\bm i}(\tau)&=&\frac{1}{\sqrt{M\beta}}\sum_{{\bm k},\omega} \tilde{\Psi}({\bm k},\omega) e^{i({\bm k}\cdot {\bm i}a-\omega\tau)}~,
\end{eqnarray}
\begin{eqnarray}
\tilde{J}'_{\bm q}
= \frac{1}{M}\sum_{\bm i} J'_{i_x} e^{i{\bm q}\cdot {\bm i}a}~,
\end{eqnarray}
\begin{eqnarray}
\varepsilon_{\bm k} = 2J\sum_{\alpha=1}^{d} \cos(k_{\alpha}a).
\end{eqnarray}
Here $M$ is the total number of sites.
In Eqs.~(\ref{eq:Psi-1}) and (\ref{eq:Psi-2}), the long-wavelength limit, $k \ll a^{-1}$, has been taken. While the terms up to the second order with respect to $ka$ is kept in the former equation that is of the order of $J$, we leave only the leading term in the latter because $J'_{i_1}$ is anticipated to be much smaller than $J$.
Substituting Eqs.~(\ref{eq:Psi-1}) and (\ref{eq:Psi-2}) into Eq.~(\ref{eq:Smid}) and taking the continuum limit $a{\bm i}\rightarrow {\bm x}$, we obtain the effective GL action,
\begin{eqnarray}
S_{\rm eff}(\{\psi\}) &&\!\!\!\!\!\!\! =\beta F_0 \!+\!
\int \!d\tau\!  \int \!d^dx\!
\left[K({\bm x}) \psi^*\frac{\partial \psi}{\partial \tau}
\!+\! W({\bm x}) \left|\frac{\partial \psi}{\partial \tau} \right|^2\right. 
\nonumber \\
&&+\Biggl.\frac{1}{2m_{\ast}}|\nabla \psi|^2+r({\bm x})|\psi|^2+\frac{u({\bm x})}{2}|\psi|^4 \Biggr],
\label{eq:Seff}
\end{eqnarray}
where the coefficients are given by
\begin{eqnarray}
K({\bm x}) &=& (ZJ)^2\beta^{(2)}({\bm x}), \label{eq:Kx} \\
W({\bm x}) &=& (ZJ)^2\gamma^{(2)}({\bm x}), \label{eq:Wx} \\
r({\bm x}) &=& ZJ + (ZJ)^2\alpha^{(2)}({\bm x}) - 2J'(x), \label{eq:rx} \\
m_{\ast} &=& \frac{1}{2Ja^2}, \label{eq:mast} \\
u({\bm x}) &=& 2a^d (ZJ)^4\alpha^{(4)}({\bm x}) \label{eq:ux}.
\end{eqnarray}
In Eq.~(\ref{eq:Seff}), we have expressed the order parameter in the dimension of the wave function as $\psi \equiv \Psi/(a^{d/2}ZJ)$. 

We assume that the system has the particle-hole symmetry when there is no potential barrier, i.e., $K_0\equiv K|_{\mu_{\bm i}=\mu_0}=0$. In this case, the potential barrier in the chemical potential $V({\bm x})$ gives the leading contribution to $K({\bm x})$ as
\begin{eqnarray}
K({\bm x}) &\simeq& K_0 - \left.\frac{\partial K}{\partial \mu} \right|_{\mu = \mu_0} V({\bm x}) = - 2W_0 V({\bm x}) \nonumber \\
&\equiv& v_{K}({\bm x}),
\label{eq:Kx_2}
\end{eqnarray}
In contrast, the contribution of $V({\bm x})$ can be ignored in the coefficients $W({\bm x})$ and $u({\bm x})$ as long as $V({\bm x}) \ll U$, and we take $W({\bm x})\simeq W|_{\mu_{\bm i}=\mu_0}\equiv W_0$ and $u({\bm x})\simeq u|_{\mu_{\bm i}=\mu_0}\equiv u_0$. The inhomogeneous hopping affects only the coefficient $r({\bm x})$,
\begin{eqnarray}
r({\bm x}) &\simeq& 
r_0 - \left.\frac{\partial r}{\partial \mu}\right|_{\mu=\mu_0} V({\bm x}) -2J'(x) 
\nonumber \\
&=& r_0 - 2J'(x) \equiv r_0 + v_{r}(x).
\label{eq:rx_2}
\end{eqnarray}
In Eq.~(\ref{eq:rx_2}), the term including $V({\bm x})$ vanishes due to the particle-hole symmetry ($K_0 = 0$). Notice that in Eqs.~(\ref{eq:Kx_2}) and (\ref{eq:rx_2}) we have used the following relations:
\begin{eqnarray}
K=-\frac{\partial r}{\partial \mu},\\
W=\frac{1}{2}\frac{\partial K}{\partial \mu},
\end{eqnarray}
which stem from the U(1) gauge invariance of the system with respect to the transformation $\psi \rightarrow \psi e^{i\phi}$ and $\mu \rightarrow \mu + i \frac{\partial \phi}{\partial \tau}$~\cite{sachdev-11}. Thus, the potential barrier created by the inhomogeneous chemical potential leads to $v_K({\bm x})$ while that by the inhomogeneous hopping leads to $v_r(x)$. This is consistent with the fact that $V_{\bm i}$ acts differently for a particle and a hole while $J'_{i_1}$ does not break the particle-hole symmetry. 
Finally, replacing the imaginary time $\tau$ with the real time $t$ as $t=-i\tau$ and minimizing the effective action, we obtain the TDGL equation Eq.~(\ref{eq:tdgl_b}). 

\begin{figure}[t]
\centerline{\includegraphics[scale=0.42]{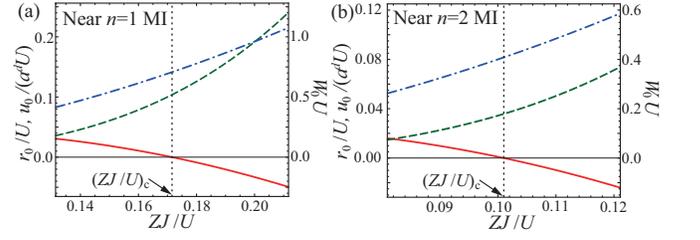}}
 \caption{(Color online) The red solid, green dashed, and blue dash-dotted lines represent the coefficients $r_0$, $u_0$, and $W_0$ in the TDGL equation as functions of $ZJ/U$ along the particle-hole symmetric line, where $K_0=0$, near the tip of the Mott insulating region with $n=1$ (a) and $n=2$ (b). The dotted lines mark the critical point $(ZJ/U)_{c}$.}
 \label{fig:GLpara}
\end{figure}
\begin{figure}[t]
\centerline{\includegraphics[scale=0.45]{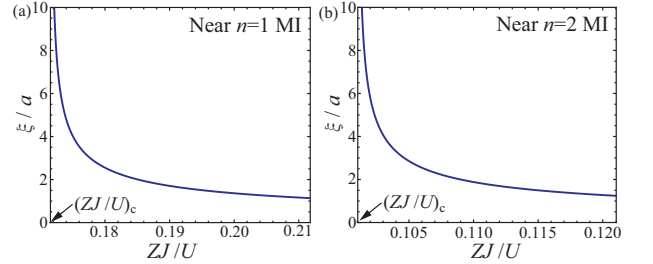}}
 \caption{Healing length $\xi$ as a function of $ZJ/U$ along the line of $K_0=0$ near the tip of the Mott insulating region with $n=1$ (a) and $n=2$ (b).}
 \label{fig:xi}
\end{figure}
\begin{figure}[t]
\centerline{\includegraphics[scale=0.45]{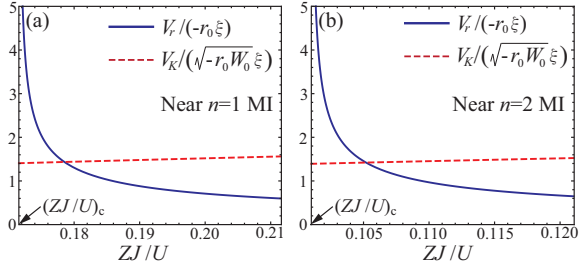}}
 \caption{(Color online) The blue solid and red dashed lines represent the strength of the potential barriers $V_r$ and $V_K$ in units of $-r_0\xi$ and $\sqrt{-r_0W_0}\xi$ along the line of $K_0=0$ near the tip of the Mott insulating region with $n=1$ (a) and $n=2$ (b). We set $J'=-0.5J$ and $V=-0.2U$.}
 \label{fig:VrVK}
\end{figure}
It is informative to evaluate the coefficients in the TDGL equation in specific parameter regions of the Bose-Hubbard model by using the analytical expressions in Eqs.~(\ref{eq:Kx})--(\ref{eq:ux}). In Fig.~\ref{fig:GLpara}, we show the coefficients $r_0$, $u_0$, and $W_0$ along the particle-hole symmetric line ($K_0=0$) as functions of $ZJ/U$. Further, the healing length $\xi=1/\sqrt{-r_0m_{\ast}}=a\sqrt{2J/(-r_0)}$ is plotted as a function of $ZJ/U$ in Fig.~\ref{fig:xi}. It diverges at the critical point $ZJ/U=(ZJ/U)_{\rm c}$ as $\xi\sim \left(ZJ/U-(ZJ/U)_{\rm c}\right)^{-1/2}$ because $r_0$ linearly vanishes near the critical point.

In order to evaluate the strength of the potential barriers $V_r$ and $V_K$, we assume a specific form of them,
\begin{eqnarray}
J_{i_1}' = J' \delta_{i_1,0}, \\
V_{i_1} = V \delta_{i_1,0}. \label{eq:Vi1}
\end{eqnarray}
Taking the continuum and thin-barrier limits leads to $V_r = -2J'a$ and $V_K = -2W_0Va$. In Fig.~\ref{fig:VrVK}, we show the dimensionless barrier strength $\tilde{V}_r\equiv V_r/(-r_0\xi)$ and $\tilde{V}_K\equiv V_K/(\sqrt{-r_0 W_0}\xi)$ as functions of $ZJ/U$. Since the former diverges as $\tilde{V}_r \sim \left(ZJ/U-(ZJ/U)_{c}\right)^{-1/2}$, in principle, one can increase it unlimitedly by approaching the critical point. In contrast, $\tilde{V}_K$
reaches a finite value at the critical point, and at a glance it seems that a certain upper limit is set by the condition $V_{i_1}\ll U$. However, this limitation stems from the choice of the single-site barrier Eq.~(\ref{eq:Vi1}). In other words, one can increase $\tilde{V}_K$ with no limit by increasing the width of the barrier. Notice that the condition that the width is much smaller than the healing length $\xi$ can be satisfied by approaching the critical point where $\xi$ diverges.

In Fig.~\ref{fig:Tandf.eps} of Sec.~\ref{sec:Tunneling}, we have taken $(\tilde{V}_r, \tilde{V}_K)=(1,1)$ and $(4,4)$ to demonstrate the Fano resonance of the NG modes through the Higgs bound state. In order to show how close to the critical point $(ZJ/U)_{\rm c}$ the system has to be for obtaining these values of $\tilde{V}_r$, i.e., how precisely the lattice depth $V_0$ has to be controlled in experiments, we specifically assume that an ultracold gas of $^{87}$Rb in the hyperfine state $|F=1,m_F=1\rangle$, whose $s$-wave scattering is given by $a_s=5.31\,{\rm nm}$, is confined in a 3D optical lattice with $a=532\,{\rm nm}$. To calculate the hopping and the onsite interaction from the experimental parameters, we use the following formulas~\cite{rey-05},
\begin{eqnarray}
J&=& A_J E_{\rm R}\left(\frac{V_0}{E_{\rm R}}\right)^{B_J}
\exp\left[-C_J\left(\frac{V_0}{E_{\rm R}}\right)^{1/2}\right], \\
U&=& \sqrt{8\pi} \frac{a_s}{a} E_{\rm R}\left(\frac{V_0}{E_{\rm R}}\right)^{3/4},
\end{eqnarray}
where $E_{\rm R}\equiv \frac{\pi^2}{2ma^2}$ denotes the recoil energy and $(A_J,B_J,C_J)=(1.397,1.051,2.121)$ are numerically obtained constants. When we take $J'=-0.5J$ as in Fig.~\ref{fig:VrVK},  $\tilde{V}_r=1$ is converted to $ZJ/U-(ZJ/U)_{c} \simeq 0.014$. This implies that the lattice depth has to be as close to the critical point as $V_{0,{c}}-V_0\simeq 0.29 E_{\rm R}$, where $V_{0,{\rm c}}$ denotes the critical lattice depth. This level of controllability has been achieved in recent experiments~\cite{endres-12}. On the other hand, $\tilde{V}_r=4$ corresponds to $ZJ/U-(ZJ/U)_{c} \simeq 0.00089$, meaning $V_{0,{c}}-V_0 \simeq 0.019 E_{\rm R}$. Such fine tuning is rather difficult even in current experiments.

\section{Solutions in the presence of a static kink condensate}
\label{sec:kinksolution}
We examine solutions of Eqs.~(\ref{eq:S(x)}) and (\ref{eq:T(x)}) without a potential barrier [$v_r(x)=v_K(x)=0$] when the
background static condensate has a kink solution $\psi_0(x)=\tanh(x)$~\cite{pitaevskiistringari-03}.  
Since the static condensate in Eq.~(\ref{eq:staticpsi0}) at $x\neq
0$ is identical to the shifted kink, solutions of $S(x)$ and
$T(x)$ for $\delta$-function potential barriers $v_r(x)=V_r\delta(x)$ and
$v_K(x)=V_K\delta(x)$ at $x\neq 0$ can be composed of those on a static kink condensate.
\par
Scattering states of $S(x)$ and $T(x)$ in the presence of a static kink are given by \cite{kovrizhin-01,Lamb}
\begin{eqnarray}
&&S(x)=\left(\tanh x-ik\right)e^{ikx}~,
\label{eq:solutionS}\\
&&T(x)=\left(3\tanh^2 x-3ik'\tanh x-(k')^2-1\right)e^{ik'x}~, \nonumber \\
\label{eq:solutionT} 
\end{eqnarray}
where
\begin{eqnarray}
\omega^2=\frac{1}{2}k^2=\frac{1}{2}k'^2+2~.
\end{eqnarray}
Each of Eqs.~(\ref{eq:solutionS}) and (\ref{eq:solutionT}) is a single plane 
wave propagating without reflection. These solutions show that the NG and Higgs modes are
not scattered by a kink, though their amplitudes are suppressed near the
kink.
\par
We note that there also exist bound-state solutions of $S(x)$ and $T(x)$ localized around
a kink. The solution $S(x)=1/\cosh x$ has imaginary frequencies
$\omega=\pm i/\sqrt{2}$ that destabilize the kink. This is in sharp contrast with the
stable 1D kink solution in the GP equation~\cite{pitaevskiistringari-03}. 
The solution $T(x)=\tanh x(1-\tanh^2x)$ has the frequency $\omega=\sqrt{3/2}$. 
In addition to these solutions, there are trivial zero-mode
solutions: $S(x)=\tanh x$ and $T(x)=1/\cosh^2 x=\frac{d}{dx}(\tanh x)$
with $\omega=0$.
\par
The solution (\ref{eq:solutionT}) is valid for $\omega$ both above and below
the Higgs gap $\Delta=\sqrt{2}$. For Higgs mode with energy above the gap ($\omega>\Delta$), 
$k'=\pm\sqrt{2\omega^2-4}\equiv\pm k_t$ is real and Eq.~(\ref{eq:solutionT})
corresponds to a scattering state. If the energy is below the gap ($\omega<\Delta$), then
$k'=\pm i\sqrt{4-2\omega^2}\equiv\pm i\kappa_t$ and thus
Eq.~(\ref{eq:solutionT}) decays as $T(x)\propto e^{\pm\kappa_t x}$ for
$x\to\mp\infty$. The bound-state solutions of $T(x)$ on a kink can be obtained by
connecting these decaying solutions.

\section{Conservation law for collective modes}
\label{sec:conservation}
We discuss conservation law for collective modes in the effective
1D setting in Fig.~\ref{fig:schematic}(b).
One can easily prove that the Wronskian of the coupled linear equations
(\ref{eq:S(x)}) and (\ref{eq:T(x)}) defined by
\begin{eqnarray}
{\mathcal W}(\phi_1(x),\phi_2(x))=
\left|
\begin{array}{cc}
\phi_1 & \phi_2\\
\frac{d\phi_1}{dx} & \frac{d\phi_2}{dx}
\end{array}
\right|~,\label{eq:Wronskian}
\end{eqnarray}
is a constant which is independent of $x$ and thus provides a conserved
quantity. Here, we assumed that $\phi_i(x)\equiv(S_i(x),T_i(x))^T$ ($i=1,2$) are solutions for the same
energy $E$ and defined the product as
\begin{eqnarray}
 \phi_1\phi_2\equiv(S_1,T_1)
\left(
\begin{array}{c}
S_2 \\
T_2
\end{array}
\right)=S_1S_2+T_1T_2~.
\end{eqnarray} 
\par
If we take 
\begin{eqnarray}
\phi_1=\left(
\begin{array}{c}
S(x) \\
T(x)
\end{array}
\right)~, \ \ \phi_2=\phi_1^*~,
\end{eqnarray}
and substitute the asymptotic forms Eqs.~(\ref{eq:asymptoticS}) and
(\ref{eq:asymptoticT}) into Eq.~(\ref{eq:Wronskian}), then we obtain 
\begin{eqnarray}
\!\! {\mathcal W}(\phi_1,\phi_2) \!=\!
\left\{
\begin{array}{ll}
\!\!
2 i k_s\left(|r_{\rm ng}|^2 \!-\!1\right), & \!\! (E \!<\! \Delta)\\
\\
\displaystyle \!\! 2 i k_s\left(|r_{\rm ng}|^2 \!-\!1\right) 
\!+\! 2 i k_t\left|r_{\rm h}\right|^2, & \!\! (E \!>\! \Delta)
\end{array} \right.\!\! ,
\label{eq:wronskian1}
\end{eqnarray}
for $x\to-\infty$, and 
\begin{eqnarray}
{\mathcal W}(\phi_1,\phi_2) \!=\!
 \left\{
\begin{array}{ll}
\! -2 i k_s |t_{\rm ng}|^2, & (E<\Delta)\\
\\
\displaystyle \! - 2 i k_s|t_{\rm ng}|^2 - 2 i k_t \left| t_{\rm h}\right|^2, &(E>\Delta)
\end{array} \right. \!\! ,
\label{eq:wronskian2}
\end{eqnarray}
for $x\to\infty$. Note that we substituted
$\kappa_t=-ik_t=-i\sqrt{2E^2-4}$ in Eq.~(\ref{eq:asymptoticT}) for $E>\Delta$. Since $\mathcal W$ is a constant,
Eqs.~(\ref{eq:wronskian1}) and (\ref{eq:wronskian2}) give the conservation of probability for the NG mode:
$|r_{\rm ng}|^2+|t_{\rm ng}|^2=\mathcal{R}+\mathcal{T}=1$ for $E<\Delta$.
For $E>\Delta$, incident the NG mode could induce Higgs mode due to the
coupling of phase and amplitude fluctuations introduced by $v_K$. If we
define the probability of the Higgs mode reflected to the left of the
potential barriers $\mathcal{R}_{\rm h}$ and emitted to the right of
the barriers $\mathcal{T}_{\rm h}$ from Eqs.~(\ref{eq:asymptoticS})
and (\ref{eq:asymptoticT}) to be 
\begin{eqnarray}
&&\mathcal{R}_{\rm h}\equiv \frac{k_t}{k_s} |r_{\rm h}|^2, \\
&& \mathcal{T}_{\rm h}
\equiv \frac{k_t}{k_s} |t_{\rm h}|^2.
\end{eqnarray}
then Eqs.~(\ref{eq:wronskian1}) and (\ref{eq:wronskian2}) give the
conservation of the total probability, including generated the Higgs mode for $E>\Delta$:
$\mathcal{R}+\mathcal{T}+\mathcal{R}_{\rm h}+\mathcal{T}_{\rm h}=1$.

\section{Decaying Higgs bound states}\label{sec:DecayingBS}

The Higgs bound states in the case $v_K(x)=0$ discussed in Sec.~\ref{sec:HiggsBS}
are eigenstates of Eq.~(\ref{eq:T(x)}) and therefore long-lived excitations within the linearized TDGL equation.
As discussed in Sec.~\ref{sec:Tunneling}, however, finite $v_K$ couples the Higgs bound states with NG mode and induces
Fano resonance of NG mode transmitting through the potential barrier. This indicates that the Higgs
bound states in the case $v_K(x)\neq 0$ are not exact eigenstates of
Eqs.~(\ref{eq:S(x)}) and (\ref{eq:T(x)}) but {\it resonant states} that
are decaying by leaking out of the double-well potential through the outgoing NG
mode. In this Appendix, we directly show that the Higgs bound states
become resonant states when $v_K\neq 0$ and estimate their lifetime.
\par
We investigate solutions of $S(x)$ and $T(x)$ when $v_r(x)=V_r\delta(x)$ and $v_K(x)=V_K\delta(x)$ in the following form: 
\begin{eqnarray}
S(x<0) &=& A_S \frac{\psi_0-ik_s}{1-ik_s} e^{- i k_s x}, \label{S(x<0)} \\
S(x>0) &=& B_S \frac{\psi_0-ik_s}{1-ik_s} e^{i k_s x}, \label{S(x>0)} \\
T(x<0) &=& A_T\frac{3\psi_0^2 + 3 \kappa_t \psi_0 + \kappa_t^2 -1}{2+3\kappa_t+\kappa_t^2}e^{\kappa_t x}, 
\label{T(x<0)}\\
T(x>0)&=&B_T \frac{3\psi_0^2 + 3 \kappa_t \psi_0 + \kappa_t^2 -1}{2+3\kappa_t+\kappa_t^2}e^{-\kappa_t x},
\label{T(x>0)}
\end{eqnarray}
The set of the boundary conditions of outgoing waves in the above equations is often called the Siegert condition and has been employed in the study of open systems such as nuclear reactions and quantum dots \cite{siegert-39,landaulifshitz-77,hatano-08}. 
\par
The coefficients $A_S$, $B_S$, $A_T$, and $B_T$ are determined so as to
satisfy the boundary conditions
Eqs.~(\ref{eq:S(0)})$\sim$(\ref{eq:T'(0)}). The condition for
nontrivial solutions to exist is given by
\begin{eqnarray}
\left|\begin{array}{cc}
k_s\left(k_s+\frac{i}{\eta}\right) & EV_Kc_2 \\
EV_K(-\eta+ik_s)  &c_1+V_rc_2
\end{array}\right|
=0. \label{eq:ASAT}
\end{eqnarray}
Setting $V_K=0$, Eq.~(\ref{eq:ASAT}) reduces to the condition for the even-parity
bound state (\ref{eq:evenparity}). We note that the odd-parity bound state remains
a long-lived eigenstate that is decoupled from NG mode because of its node at $x=0$.
To see effects of $v_K(x)$ on the even-parity bound state, we expand
Eq.~(\ref{eq:ASAT}) by $V_K$ and furthermore use the expansion around
$E_+$ in Eq.~(\ref{eq:expansionE+}). We thus obtain the energy of the eigenstate: 
\begin{eqnarray}
 E \!=\! E_{+}
\!-\!
\frac{E_{+} c_2 \left(\left(\frac{1}{\eta} - \eta\right)\sqrt{2}E_{+} 
\!+\! i(2E_{+}^2 \!+\! 1)\right)}
{\sqrt{2} \alpha \left(2E_{+}^2 + \frac{1}{\eta^2}\right)}V_K^2.
\label{eq:Eresonant}
\end{eqnarray}
The negative imaginary part in Eq.~(\ref{eq:Eresonant}) clearly shows
that the eigenstate decays exponentially in time, and therefore the even-parity
bound state becomes a resonant state in the presence of $v_K(x)$.
The lifetime of the resonant state is proportional to the inverse of the
imaginary part and of the order of $1/V_K^2$. 

\section{Spectral function of the Higgs bound states}\label{sec:spectral}
In Sec.~\ref{sec:QFT}, we have formulated quantum field theory for the effective action of Eq.~(\ref{eq:Seff}), which possesses both first- and second-order time-derivative terms, in terms of the quasi-particle basis. In this Appendix, we use the formulation within the quadratic approximation to calculate the spectral function of the Higgs bound states in the presence of finite $K({\bm x})$, which breaks the particle-hole symmetry. In the case of the local potential barrier of $K({\bm x})=V_K\delta(x)$, we will show that the decay rate defined as the peak width of the spectral function precisely agrees with that calculated by using the Siegert boundary condition in Appendix~\ref{sec:DecayingBS}. We also consider a global shift of the potential, i.e., $K({\bm x})=K_0 \neq 0$, to reveal that the binding energy monotonically increases when $K_0$ increases.

We aim to calculate the Green's function of the Higgs bound states within the quadratic approximation,
\begin{eqnarray}
G_{l'}'(i\omega_n) = - \langle \alpha_{l'}'(\omega_n)
\left(\alpha_{l'}'(\omega_n)\right)^{\ast} \rangle_0,
\label{eq:green_def}
\end{eqnarray}
and the spectral function,
\begin{eqnarray}
\rho_{l'}'(\omega) =
-2\,{\rm Im}\left[ G_{l'}'(i\omega_n\rightarrow\omega + i0^{+}) \right],
\label{eq:rho_def}
\end{eqnarray}
where $\alpha_{l'}'(\omega_n)$ denotes the quasi-particle field of the Higgs bound states for $K({\bm x})=0$ and the index $l'\in \{{\rm e},{\rm o}\}$ specifies the even- or odd-parity bound state. The average $\langle \cdots \rangle_0$ is taken with the quadratic action for $K({\bm x})\neq 0$. As mentioned in Appendix~\ref{sec:DecayingBS}, while the Higgs bound states localized around the barriers are eigenstates with infinite lifetime in the case of $K({\bm x})=0$, finite $K({\bm x})$ forces the bound states to decay into the NG modes. The peak width of the spectral function of Eq.~(\ref{eq:rho_def}) is interpreted as the decay rate.

In order to calculate the Green's function of Eq.~(\ref{eq:green_def}), one needs to relate ${\boldsymbol \alpha}_{l'}'(\omega_n)\equiv \left[\alpha_{l'}'(\omega_n), (\alpha_{l'}'(-\omega_n))^{\ast}\right]^{\bf t}$ to ${\boldsymbol \alpha}_{l}(\omega_n)$. Using the inverse transformation of Eq.~(\ref{eq:inverse}) at $K({\bm x})=0$, we relate ${\boldsymbol \alpha}_{l'}'$ to $\tilde{\boldsymbol \chi}$,
\begin{eqnarray}
{\boldsymbol \alpha}_{l'}' = 
 \int d{\bf x} \, i\hat{\sigma}_z (\hat{X}_{l'}')^{\dagger}\hat{Q}' \tilde{\bm \chi}
\label{eq:inverse_dash}
\end{eqnarray}
where $\hat{X}_{l'}'$ and $\hat{Q}'$ means $\hat{X}_{l}$ of Eq.~(\ref{eq:Xmat}) and $\hat{Q}$ of Eq.~(\ref{eq:Qmat}) for $K=0$, namely,
\begin{eqnarray}
\hat{X}_{l'}'({\bm x}) =
\left[
\begin{array}{cc}
\eta_{\sigma,l'}'({\bm x}) & \left(\eta_{\sigma,l'}'({\bm x})\right)^{\ast} 
\\
\zeta_{\sigma,l'}'({\bm x}) & \left(\zeta_{\sigma,l'}'({\bm x}) \right)^{\ast}
\\
\eta_{\varphi,l'}'({\bm x}) & \left(\eta_{\varphi,l'}'({\bm x}) \right)^{\ast}  
\\  
\zeta_{\varphi,l'}'({\bm x}) & \left( \zeta_{\varphi,l'}'({\bm x}) \right)^{\ast}
\end{array}
\right],
\end{eqnarray}
and
\begin{eqnarray}
\hat{Q}' =
\left[
\begin{array}{cccc}
0 & 1 & 0 & 0 \\
-1 & 0 & 0 & 0 \\
0 & 0 & 0 & 1 \\
0 & 0 & -1 & 0 
\end{array}
\right].
\end{eqnarray}
Here 
$
{\bm y}_{l'}'  \equiv 
\left[ \eta_{\sigma,l'}', \zeta_{\sigma,l'}', \eta_{\varphi,l'}', \zeta_{\varphi,l'}' \right]^{\bf t}
$
and $E_{l'}'$ denote the solution of Eq.~(\ref{eq:bogolike}) for $K=0$. Notice that since the phase and amplitude modes are completely decoupled at $K=0$, $\eta_{\varphi, l'}'=\zeta_{\varphi, l'}'=0$ for an amplitude mode and $\eta_{\sigma, l'}'=\zeta_{\sigma, l'}'=0$ for a phase mode. Combining Eqs.~(\ref{eq:cano}) and (\ref{eq:inverse_dash}), we express the linear transformation between ${\boldsymbol \alpha}_{l'}'$ and ${\boldsymbol \alpha}_{l}$,
\begin{eqnarray}
{\boldsymbol \alpha}_{l'}' &=& \sum_{l}\int d{\bm x}\, i\hat{\sigma}_z
(\hat{X}_{l'}')^{\dagger}\hat{Q}\hat{X}_l {\boldsymbol \alpha}_{l}
\nonumber \\
&=& \left[
\begin{array}{cc}
\mathcal{L}_{l'}(E_l) & \mathcal{M}^{\ast}_{l'}(E_l) \\
\mathcal{M}_{l'}(E_l) & \mathcal{L}_{l'}^{\ast}(E_l)
\end{array}
\right]
{\boldsymbol \alpha}_{l},
\label{eq:linear_dash}
\end{eqnarray}
where
\begin{eqnarray}
&&\!\!\!\!\!\!\!\!\!\!\!\!\!\!\!\! \mathcal{L}_{l'}(E_l) =  \int d^d x \, i \sum_{\chi} 
\left( 
(\eta_{\chi,l'}')^{\ast}\zeta_{\chi,l} - (\zeta_{\chi,l'}')^{\ast}\eta_{\chi,l}
\right),
\\
&&\!\!\!\!\!\!\!\!\!\!\!\!\!\!\!\! \mathcal{M}_{l'}(E_l) = \int d^d x \, i \sum_{\chi} 
\left( 
\eta_{\chi,l'}' \zeta_{\chi,l} - \zeta_{\chi,l'}' \eta_{\chi,l}
\right).
\end{eqnarray}

Substituting Eq.~(\ref{eq:linear_dash}) into Eq.~(\ref{eq:green_def}), we obtain the Green's function,
\begin{eqnarray}
G_{l'}'(i\omega_n) 
= \sum_{l}\left[
\frac{|\mathcal{L}_{l'}(E_l)|^2}{i\omega_n - E_l} 
+ \frac{|\mathcal{M}_{l'}(E_l)|^2}{-i\omega_n - E_l}
\right],
\end{eqnarray}
and the spectral function,
\begin{eqnarray}
\rho_{l'}'(\omega)
\!\!&=& \!\! 2\pi \sum_l \left[ 
|\mathcal{L}_{l'}(E_l)|^2\delta\left(\omega - E_l \right)
\right.
\nonumber \\
&&\left. 
- |\mathcal{M}_{l'}(E_l)|^2\delta\left(\omega + E_l \right) 
\right].
\label{eq:rho_mid}
\end{eqnarray}
Equation (\ref{eq:rho_mid}) tells that the overlap integrals $\mathcal{L}_{l'}(E_l)$ and $\mathcal{M}_{l'}(E_l)$ determine the spectral function. In order to evaluate them, one needs to be aware of the solutions ${\bm y}_{l'}'$ and ${\bm y}_l$ of the linear equation (\ref{eq:bogolike}). On the one hand, the Higgs bound-state solution ${\bm y}_{l'}'$ for $K({\bm x})=0$ is given by the product of the solution of Eq.~(\ref{eq:HBS_sol}) and the normalization factor as
\begin{eqnarray}
\eta_{\sigma,l'}' = \frac{1}{\mathcal{N}_{l'}'}T(x), \,\, \eta_{\varphi,l'}'=0.
\label{eq:bound_sigma}
\end{eqnarray}
On the other hand, the solution ${\bm y}_l$ for $v_K(x)\neq 0$ is given by
\begin{eqnarray}
\!\!\!\!\! \eta^{({e})}_{\sigma,k_s}\!\!\! &=& \!\!\!
\frac{2}{\mathcal{N}_{\rm e}}
t_{\rm h}
\frac{3\psi_0^2+3\kappa_t \psi_0+\kappa_t^2-1}{2+3\kappa_t+\kappa_t^2}
e^{-\kappa_t |x|},
\label{eq:even_sigma}
\\
\eta^{({e})}_{\varphi,k_s}\!\!\!&=& \!\!\!
-\frac{i}{\mathcal{N}_{\rm e}}\left(
\frac{\psi_0+ik_s}{1+ik_s}e^{-i k_s |x|}
\right.
\nonumber \\
&& \,\,\,\,\, \left. + (r_{\rm ng} + t_{\rm ng})\frac{\psi_0-ik_s}{1-ik_s}e^{ik_s|x|}
\right),
\label{eq:even_phi}
\end{eqnarray}
and
\begin{eqnarray}
\!\!\!\!\! \eta^{({o})}_{\sigma,k_s} &=& 
0,
\label{eq:odd_sigma}
\\
\eta^{({o})}_{\varphi,k_s} &=&
- \frac{i\,{\rm sgn}(x)}{\mathcal{N}_{\rm o}}\left(
\frac{\psi_0+ik_s}{1+ik_s}e^{-i k_s |x|}
\right.
\nonumber \\
&& \,\,\,\,\, \left. + (r_{\rm ng} - t_{\rm ng})\frac{\psi_0-ik_s}{1-ik_s}e^{ik|x|}
\right),
\label{eq:odd_phi}
\end{eqnarray}
where $\mathcal{N}_{e/o}$ is the normalization constant determined by Eq.~(\ref{eq:ortho1}) and we used the relation $t_{\rm h}=r_{\rm h}$. The solution of Eqs.~(\ref{eq:even_sigma}) and (\ref{eq:even_phi}) has even parity while that of Eqs.~(\ref{eq:odd_sigma}) and (\ref{eq:odd_phi}) has odd parity. The solution with even (odd) parity  is constructed by the summation (subtraction) of the left-incident solution of Eqs.~(\ref{eq:tunnelS(x)}) and (\ref{eq:tunnelT(x)}) and the right-incident solution, which can be obtained in a similar way. The odd-parity solution does not contribute at all to the spectral function of the Higgs bound states, because its amplitude sector is zero, as shown in Eq.~(\ref{eq:odd_sigma}). 

\begin{figure}[t]
\centerline{\includegraphics[scale=0.4]{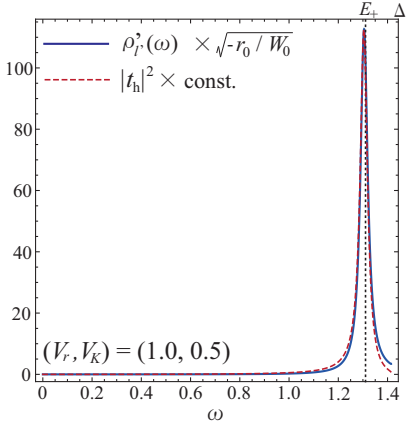}}
 \caption{(Color online) The blue solid line represents the spectral function $\rho_{l'}'(\omega)$ for the even-parity Higgs bound state at $(V_r,V_K)=(1.0,0.5)$. The red dashed line represents $|t_{\rm h}|^2\times {\rm const.}$, where the constant is determined such that the peak height matches that of $\rho_{l'}'(\omega)$. The horizontal axis is in unit of $\sqrt{-r_0/W_0}$.}
 \label{fig:rho}
\end{figure}
\begin{figure}[t]
\centerline{\includegraphics[scale=0.4]{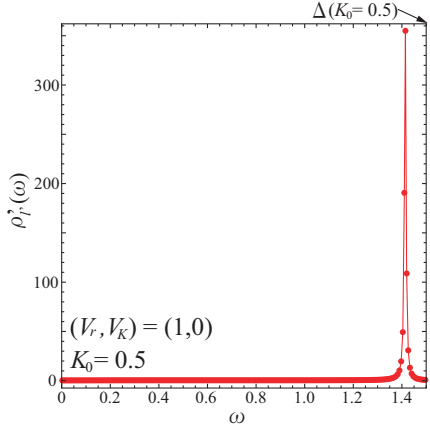}}
 \caption{(Color online) The red dots represent the spectral function $\rho_{l'}'(\omega)$ for the even-parity Higgs bound state at $(V_r,V_K)=(1,0)$, $K_0=0.5$, and $L=600$. Recall that the unit of the length is the healing length $\xi$. The red solid line is a guide to the eye. The vertical and horizontal axes are in units of $\sqrt{-W_0/r_0}$ and $\sqrt{-r_0/W_0}$.}
 \label{fig:rho2}
\end{figure}
\begin{figure}[t]
\centerline{\includegraphics[scale=0.5]{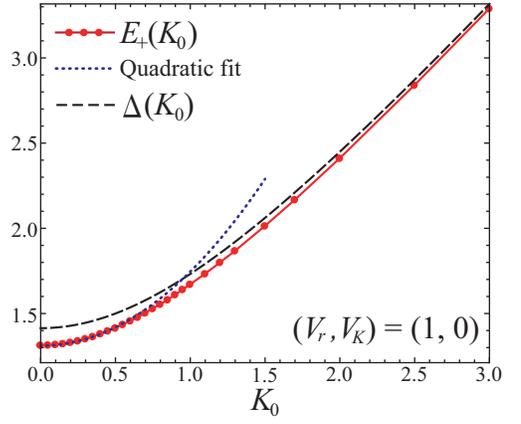}}
 \caption{(Color online) The red dots represent the binding energy $E_{+}$ as a function of $K_0$, where $(V_r,V_K)=(1,0)$ and $L=600$. The red solid line is guide to the eye.  The blue dotted line represents the best fit to the data in the region of $K_0\leq 0.5$ with use of the quadratic fitting function, $E_+(K_0)=E_+(0)+C_{\rm quad}K_0^2$, where $C_{\rm quad}$ is a free parameter. The black dashed line represents the gap energy $\Delta$ of the gapful mode in the bulk as a function of $K_0$~\cite{michikazu-15}. The vertical and horizontal axes are in units of $\sqrt{-r_0/W_0}$ and $\sqrt{-r_0 W_0}$.}
 \label{fig:Eplus}
\end{figure}

Further, the wave number $k_s$ of the NG mode is naturally chosen as the quantum number such that the summation with respect to the index $l$ in Eq.~(\ref{eq:rho_mid}) is replaced as $\sum_l = \frac{L}{2\pi}\int_0^{\infty}dk_s = \frac{L}{\sqrt{2}\pi}\int_0^{\infty}dE$, where $L$ is the system size in the $x$ direction. Carrying out the integral leads to
\begin{eqnarray}
\rho_{l'}'(\omega)\!=\! 
\sqrt{2}L\! \left[|\mathcal{L}_{l'}(\omega)|^2\theta(\omega) 
\!-\! |\mathcal{M}_{l'}(-\omega)|^2\theta(-\omega)\right]\! .
\label{eq:rho_final}
\end{eqnarray}
Substituting Eqs.~(\ref{eq:bound_sigma}) and (\ref{eq:even_sigma}) into Eq.~(\ref{eq:rho_final}), we calculate the spectral function and plot that for the even-parity Higgs bound state at $(V_r,V_K)=(1.0,0.5)$ in Fig.~\ref{fig:rho}. There we see that the spectral function has a peak with finite width near the binding energy $E_{+}$. It is remarkable that the peak shape is well approximated by the shape of $|t_{\rm h}|^2$. This means that the peak structure of the spectral function originates from $t_{\rm h}$ in Eq.~(\ref{eq:even_sigma}). Through Eq.~(\ref{eq:T'(0)}), $|t_{\rm h}|^2$ can be related to $|t_{\rm ng}|^2$, whose analytical expression is given by Eq.~(\ref{eq:transmission1}), as
\begin{eqnarray}
|t_{\rm h}|^2= |t_{\rm ng}|^2
\frac{E^2V_K^2(\eta^2+2E^2)(2+3\kappa_t+\kappa_t^2)^2}{(1+2E^2)(c_1+V_r c_2)^2}.
\label{eq:t_h^2}
\end{eqnarray}
Expanding Eq.~(\ref{eq:t_h^2}) around $E=E_{+}$ and assuming $V_K \ll 1$, we show that the peak of $|t_{\rm h}|^2$ takes a Lorentzian shape,
\begin{eqnarray}
|t_{\rm h}|^2 \propto \frac{1}
{\left(E-E_{+} - \frac{c_A c_B}{(1+c_A)\alpha}\right)^2 
+ \frac{c_A c_B^2}{(1+c_A)^2\alpha^2}}
\label{eq:lorentzian}
\end{eqnarray}
where
\begin{eqnarray}
c_A &=& \frac{2E_{+}^2 V_r^2}{(2E_{+}^2+1)^2}, \\
c_B &=& V_K^2 c_2\frac{2E_{+}^2+\eta^2}{2V_r}.
\end{eqnarray}
From Eq.~(\ref{eq:lorentzian}) it is easy to obtain the width of the Lorentzian,
\begin{eqnarray}
w = \frac{\sqrt{c_A}c_B}{(1+c_A)\alpha}
= \frac{ E_+ c_2(2E_+^2 +1)}
{\sqrt{2}\left(2E_+^2 +\frac{1}{\eta^2}\right)\alpha}V_K^2,
\end{eqnarray}
which corresponds to the decay rate of the even Higgs bound state. Indeed, this precisely agrees with the decay rate obtained in Appendix~\ref{sec:DecayingBS} through the Siegert boundary condition [see the imaginary part of Eq.~(\ref{eq:Eresonant})].

We note that the spectral function of the odd-parity Higgs bound state takes a simple $\delta$-function form,
\begin{eqnarray}
\rho_{l'={\rm o}}'(\omega) = 2\pi \delta(\omega - E_{-}).
\end{eqnarray}
This happens because the state is not coupled with the NG modes via the $v_K(x)$ potential.

We next analyze the spectral function for a homogeneous potential $K({\bm x})=K_0$. In this case, since we have not found analytical solutions of Eq.~(\ref{eq:bogolike}), we numerically solve it for a finite-sized but large system to obtain the eigenenergies $E_l$ and the eigenfunctions ${\bm y}_l({\bm x})$. Substituting the obtained solutions into Eq.~(\ref{eq:rho_final}), we calculate the spectral function of the even-parity Higgs bound state as plotted in Fig.~\ref{fig:rho2}, where $(V_r,V_K)=(1,0)$, $K_0=0.5$, and $L=600$. Notice that the spectral function does not depend on the sign of $K_0$. In Fig.~\ref{fig:rho2}, we see that the peak is slightly broadened, meaning that the lifetime of the bound state is finite. The peak position corresponding to the binding energy is shifted to the high-frequency side. In Fig.~\ref{fig:Eplus}, we show the binding energy $E_{+}$ as a function of $K_0$. In the region of $K_0<0.5$, where the bound state consists dominantly of amplitude fluctuation, $E_{+}$ increases quadratically with increasing $K_0$.  When $K_0$ increases further, the binding energy asymptotically approaches the gap energy of the gapped mode in the bulk, $\Delta(K_0)$, from its lower side. In this region of large $K_0$, the bound state is no longer a collective mode but a single-particle state, in which amplitude and phase fluctuations are substantially mixed.

\end{document}